\documentclass[10pt,journal,final,twocolumn]{IEEEtran}
\usepackage{amsmath,graphicx,subfigure, amsfonts}
\usepackage{enumerate}
\usepackage[english]{babel}
\newcommand\floor[1]{\lfloor#1\rfloor}

\ifCLASSINFOpdf
\else
\fi
\begin{document}
%
\title{On Marine Mammal Acoustic Detection Performance Bounds}
%
%
%

\author{Yin Xian,~
        Loren Nolte,~
        Stacy Tantum,~
        Xuejun Liao~
        and Yuan Zhang
\thanks{Yin Xian, Loren Nolte, Stacy Tantum and Xuejun Liao are with the Department
of Electrical and Computer Engineering, Duke University, Durham,
NC, 27708 USA e-mail: (yx15@duke.edu, lwn@ee.duke.edu).}
\thanks{Yuan Zhang is with Department of Mathematics, UCLA, Los Angeles, CA 90095, USA email:(yuanzhang@math.ucla.edu).}}
\maketitle

\begin{abstract}
Since the spectrogram does not preserve phase information contained in the original data, any algorithm based on the spectrogram is not likely to be optimum for detection. In this paper, we present the Short Time Fourier Transform detector to detect marine mammals in the time-frequency plane. The detector uses phase information for detection. We evaluate this detector by comparing it to the existing spectrogram based detectors for different SNRs and various environments including a known ocean, uncertain ocean, and mean ocean. The results show that this detector outperforms the spectrogram based detector. Simulations are presented using the polynomial phase signal model of the North Atlantic Right Whale (NARW), along with the bellhop ray tracing model.
\end{abstract}


%
\IEEEpeerreviewmaketitle

\section{Introduction}
\label{sec:intro1}
 Many marine mammal vocalizations are non-periodic and frequency modulated signals~\cite{beecher, clark1}. For this reason, the time-frequency analysis techniques are widely applied in analyzing such signals. Many time-frequency representation methods, such as the Short Time Fourier Transform (STFT), the Wigner Ville distribution and the wavelet transform have been used in this field. When we use the STFT, we can generate the spectrogram to examine the energy distribution of the signal in the time-frequency plane.  Spectrograms are among the simplest and natural tools to analyze the AM-FM type signals~\cite{flandrin}. It can sparsely represent the signals' energy ribbons localized along the time-frequency trajectories~\cite{flandrin2}. Many detection methods have been proposed to detect marine mammals based on the spectrogram, such as spectrogram correlation~\cite{mellinger1, clark3}, frequency contour edge detection method~\cite{gillespie} and contour extraction~\cite{mellinger2, buck, Madhusudhana,mallawaarachchi}. In order to obtain the optimal detection performance based on the spectrogram, it is natural to examine the probability distribution of the spectrogram data, and obtain the likelihood ratio for detection. Analysis of the probability distribution of the spectrogram elements has been performed to approximate the likelihood ratio of the spectrogram by assuming that the spectrogram elements are statistically independent~\cite{atles}. Research can be found to perform detection based on the likelihood ratio of a single spectrogram element~\cite{Martin, Martin3}.

However, the phase information is neglected when we do detection based on the spectrogram. The phase information is particularly useful for signal reconstruction and source localization~\cite{oppenheim}. Studies have found that the phase spectrum is more sensitive than the magnitude spectrum for signal recognition when the window function of the STFT is appropriately chosen~\cite{liu, paliwal}.  By incorporating the phase information and the magnitude information, we can improve the detection performance. In this paper, we propose the STFT detector. Instead of extracting the phase information from the source signal separately for detection, we incorporate the magnitude and phase information during the STFT. Results show that the STFT detector outperforms detectors based on the spectrogram, and can achieve optimal detection result when the environmental parameters and source signal are known exactly, and the noise is additive white Gaussian. Technical details of the detector can be found in Section~\ref{sec:detection_model} and the Appendix.

The sound speed in the ocean is influenced by temperature, salinity,  and pressure of the sea water, so it varies spatially~\cite{preisig, johnson1}. The speed of sound changes with depth, yielding what is known as a sound speed profile. The spatial variability gives rise to refraction of the sound. Refraction and reflection from the sea surface and bottom contribute to multipath propagation~\cite{johnson1}. Multipath ocean propagation environment will lead to dispersion and time spread for the transmitted signal~\cite{rouseff}. It is a challenge for the underwater acoustic communication systems. Research has been done to evaluate the uncertainty of environmental parameters to source localization~\cite{shorey1998wideband, book1997, stacy}, and detection in the time domain~\cite{sha, wazenski}. In this paper, we will evaluate the influence of the uncertainty of the sound speed profile on detection in the time-frequency plane. The ROC of the time domain matched filter assuming the signal is known exactly is used to benchmark the detection performance of our proposed detector.

The organization of the paper is as follows. Section~\ref{sec:intro1} gives an introduction and background for this paper. Section~\ref{sec:prop_model} presents the ocean acoustics propagation model, environmental parameters setting and propagated signal model. Overviews of the characteristics of the current detection model, and statistics of our proposed STFT detector are presented in Section~\ref{sec:detection_model}. Section~\ref{sec:result} gives detection performance results using the STFT detector in the matched ocean, uncertain ocean and mean ocean cases. Section~\ref{sec:conclusion} concludes the paper.

\medskip

\section{Ocean Acoustics Propagation Model}
\label{sec:prop_model}
\medskip

\subsection{\textbf{Environment Configuration}}
\begin{figure}[ht!]
\begin{center}
\includegraphics[height=53mm,width=80mm]{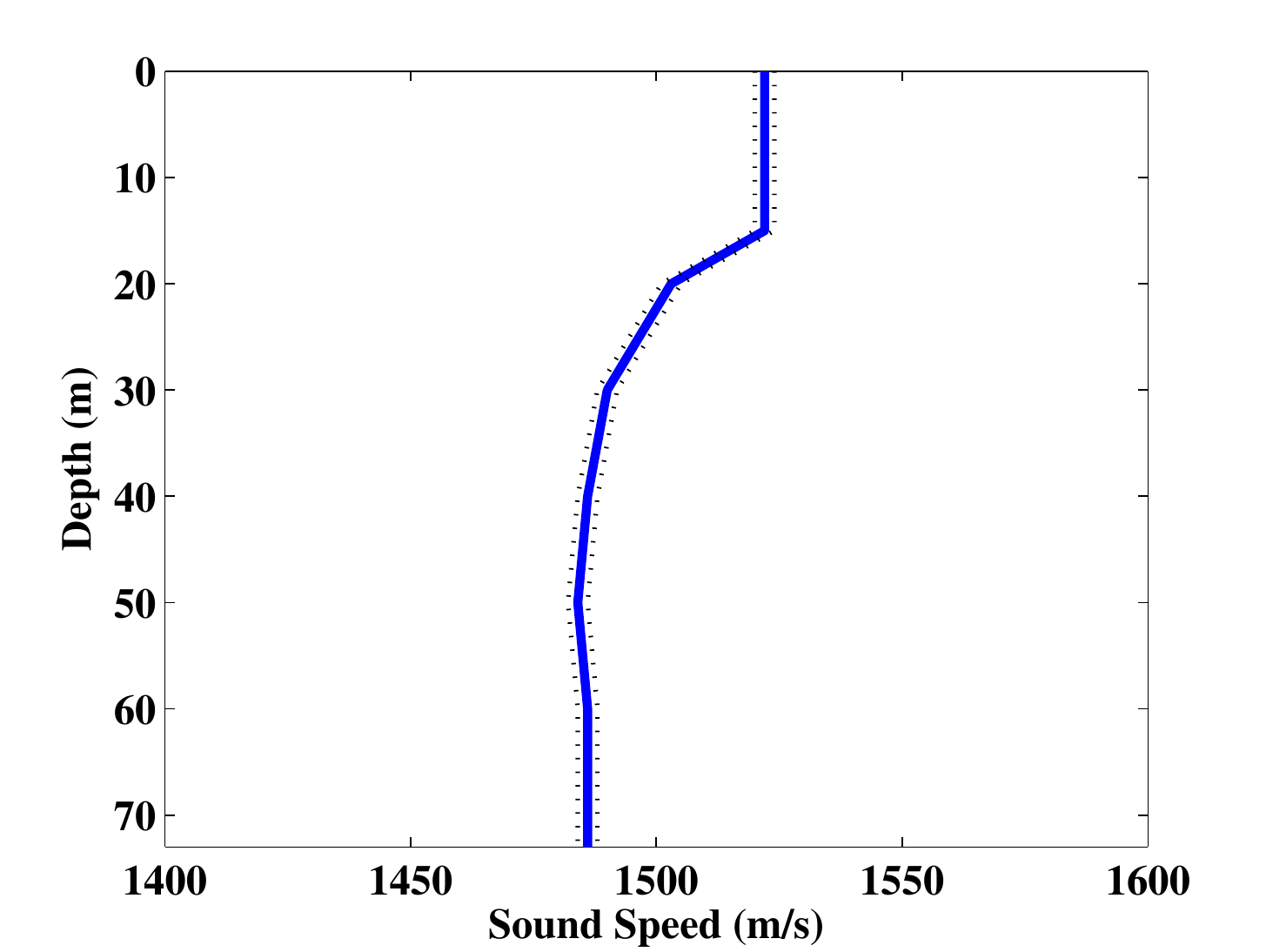}
\end{center}
\vspace{-12pt}
\caption{{Hudson Canyon Typical Sound Speed Profile}}\label{fig:ssp}
\vspace{-10pt}
\end{figure}

 We use the Hudson Canyon as the ocean propagation environment. The Hudson Canyon typical sound speed profile is shown in Figure~\ref{fig:ssp}. The Hudson Canyon environment is characterized by a sandy bottom and a sound speed profile. The environmental parameters and uncertainties of the canyon are shown in Table~\ref{table:hudson_para}~\cite{shorey1998wideband}.

\begin{table}[h!]
\footnotesize
\caption{Hudson Canyon Environment Parameters and Uncertainties} 
\label{table:hudson_para}
\centering 
\begin{tabular}{cccc}
\hline\hline
~ & ~ & Basis & Uniform \\
Description & Depth &  sound speed & uncertainty \\ [1ex]
\hline
Water column sound speed & 0+ m & 1522 m/s & $\pm$ 2m/s\\ 
~ & 15 m & 1522 m/s & $\pm$ 2m/s\\
~ & 20 m & 1503 m/s & $\pm$ 2m/s\\
~ & 30 m & 1490 m/s & $\pm$ 2m/s\\
~ & 40 m & 1486 m/s & $\pm$ 2m/s\\
~ & 50 m & 1484 m/s & $\pm$ 2m/s\\
~ & 60 m & 1486 m/s & $\pm$ 2m/s\\
~ & 73- m & 1486 m/s & $\pm$ 2m/s\\
Bottom half-space sound speed & 73+ m & 1550 m/s & $\pm$ 50m/s\\[1ex]
\hline 
\end{tabular}
\end{table}

\subsection{\textbf{Signal Model}}
Many marine mammal vocalizations are frequency modulated, and can be modelled as polynomial-phase signals~\cite{clark1}. Take the North Atlantic Right Whale (NARW) as an example. It is found that there are nine types of sound for the NARW according to its time-frequency characteristics~\cite{trygonis2013vocalization}, and the upsweep call is commonly found when the whales greet each other.

Assume that $s(t)$ is the source signal of the whale, $\theta(t)$ is the phase of the signal, $a$ is an unknown positive scalar representing the amplitude, so~\cite{clark1}
\begin{align}
s(t)=a\cos(\theta(t))=a\cos\Bigl(2\pi\sum\limits_{m=0}^{M-1}f_m t^m\Bigr),  ~t=1,2,\cdots,N, \label{eq:signal_model}
\end{align}
\normalsize
where $f_0, f_1,\cdots,f_{M-1}$ are polynomial coefficient parameters for the signal model.

\begin{figure}[ht!]
     \begin{center}
        \subfigure[Real NARW Contact Call]{%
            \includegraphics[width=.235\textwidth,height=3.6cm]{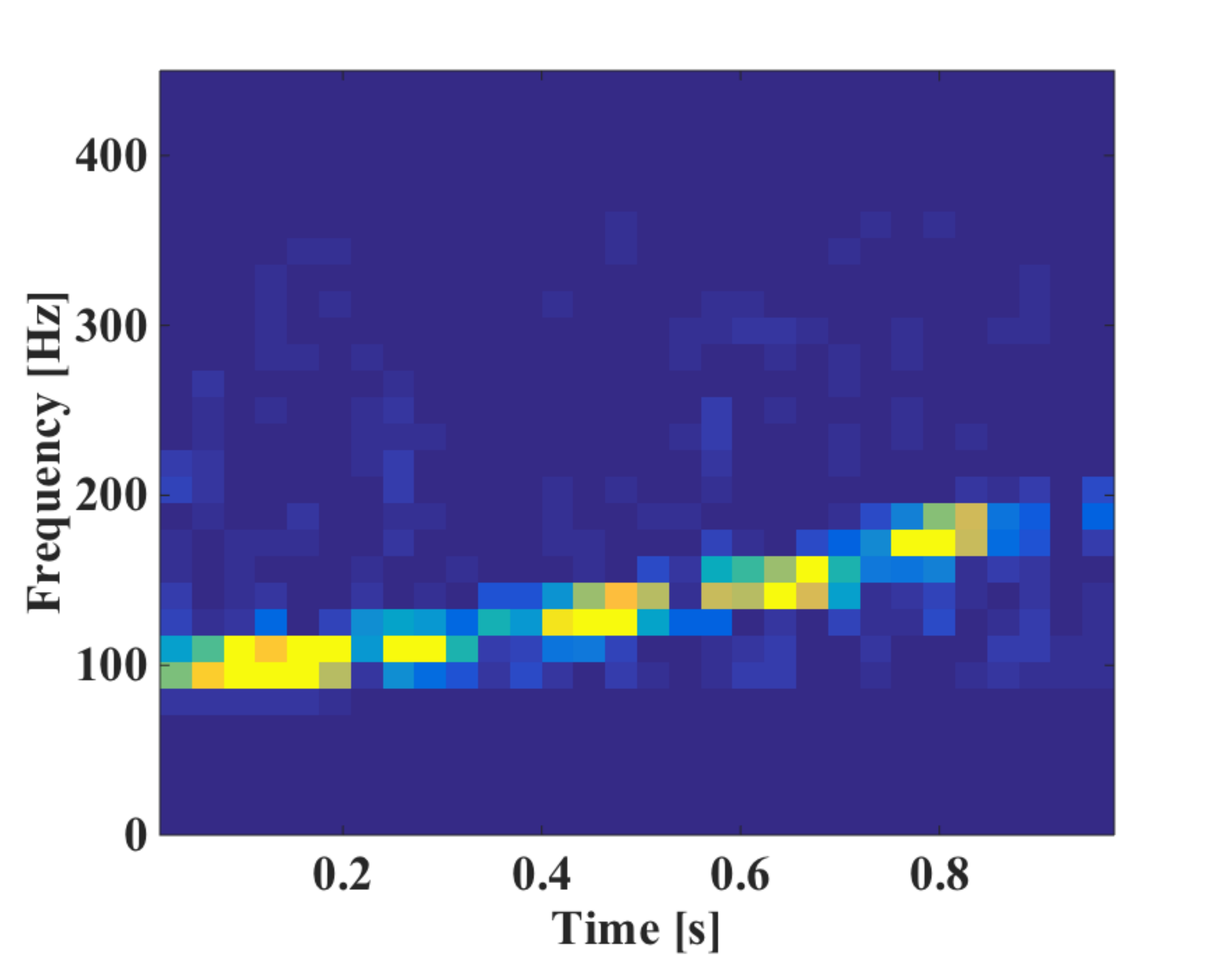}
        }%
        \subfigure[Synthetic NARW Contact Call]{%
           \includegraphics[width=.235\textwidth,height=3.6cm]{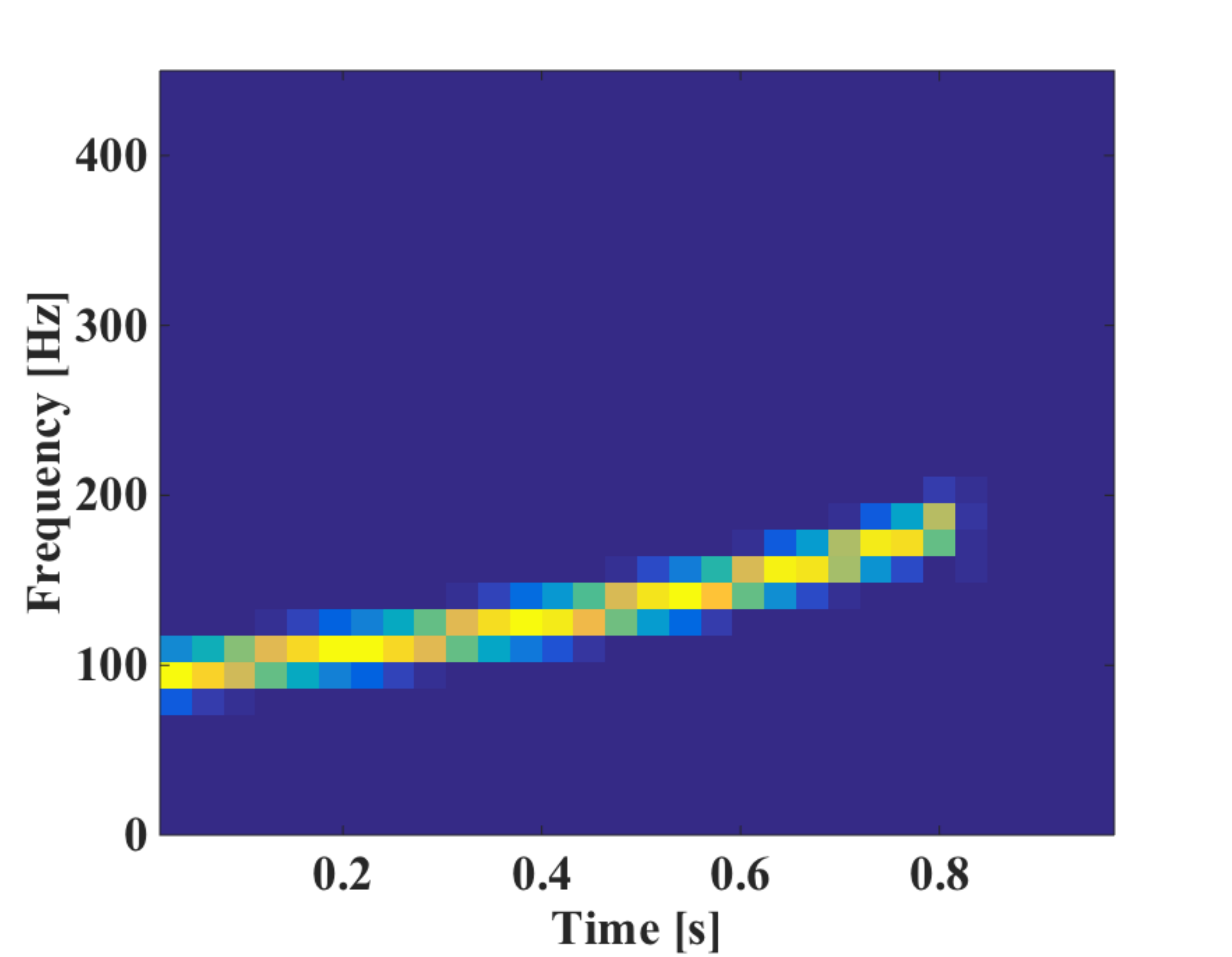}
        }
    \end{center}
    \vspace{-10pt}
    \caption{North Atlantic Right Whale Signal}\label{fig:synthetic}
    \vspace{-10pt}
\end{figure}

By inspecting the shape of the whales' sound contour, and fitting the polynomial coefficients, we can synthesize the right whale
signal (Figure~\ref{fig:synthetic}). We will use the synthetic NARW signal to evaluate the effect of propagation environment on acoustic detection performance.

\medskip

\subsection{\textbf{Signal Propagation}}
\begin{figure}[ht!]
\begin{minipage}{\linewidth}
\begin{center}
\includegraphics[height=53mm,width=80mm]{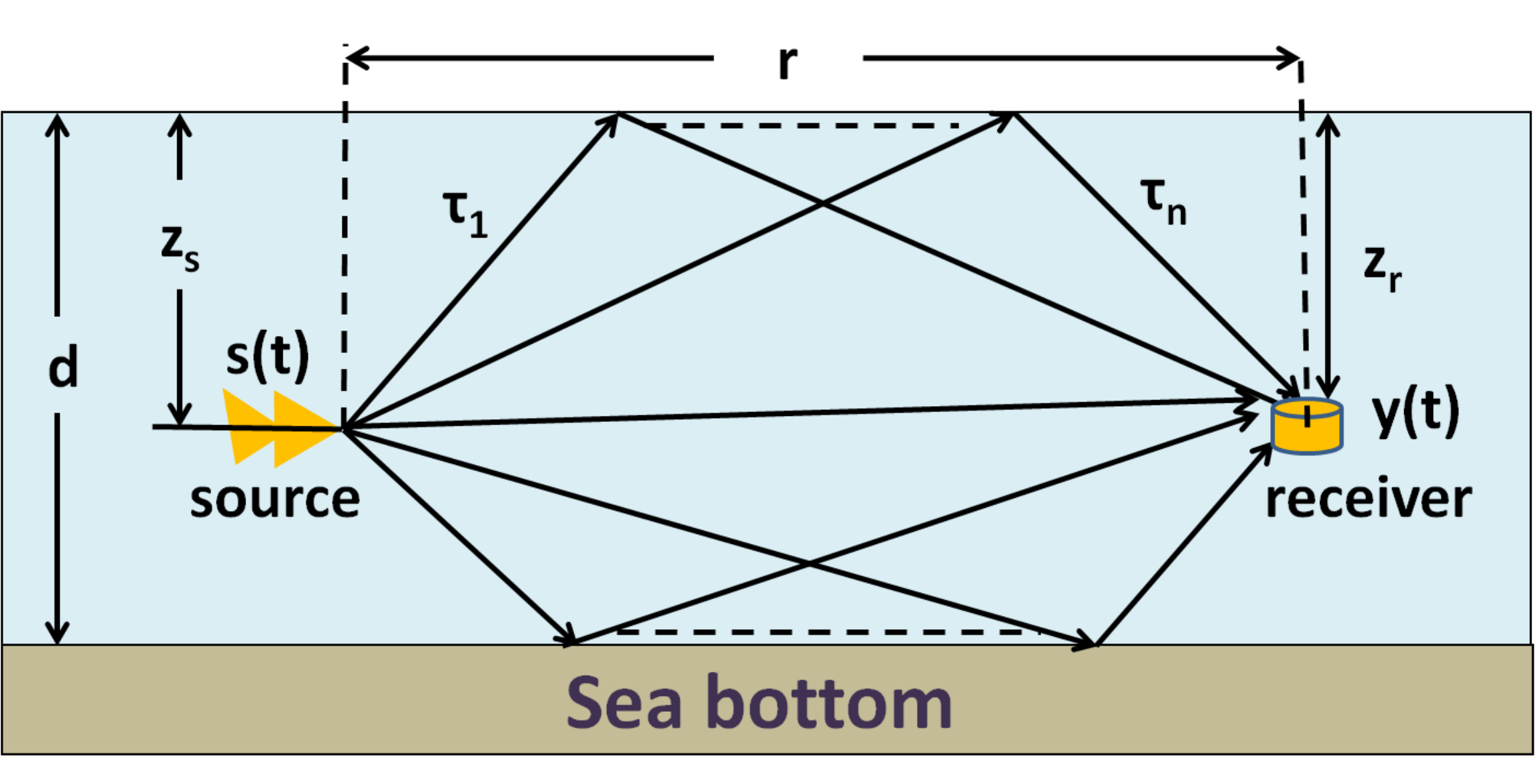}
\end{center}
\vspace{-10pt}
\caption{Ocean Acoustic Propagation Model} \label{fig:multipath}
\end{minipage}
\end{figure}

Supposed that the whale is $z_s$ meters below the sea surface. We want to determine whether there is a signal present at the receiver, which is $r$ meters from the source and $z_r$ meters below the sea surface. The environment setting can be characterized as in Figure~\ref{fig:multipath}.
\begin{table}[ht!] 
\footnotesize
\centering
\vspace{-17pt}
\begin{tabular}{l l l l}
where, &~ &~ &~\\
$z_s$: source depth; &~  &~ &$z_r$: source depth; \\
$s(t)$: source signal; &~ &~  &$y(t)$: propagated signal;  \\
$r$: range of propagation; &~ &~ &$d$: depth of shallow water; \\
$\tau_1$: delay of the first path;&~ &~ &$\tau_n$: delay of the n-th path.
\end{tabular}
\end{table}

We use the Bellhop Model~\cite{porter2011bellhop} to calculate the arrival amplitudes and delays of all possible signal paths in the ocean. Supposing there are $K$ arrivals, the propagated signal $y(t)$ can be represented as the sum of $K$ arrival signals. The amplitude and delay of each arrival signal is $A_k$ and $\tau_k$ respectively, that is
\begin{align}
y(t)=\sum\limits_{k=1}^{K}A_{k}s(t-\tau_k)
\end{align}
where $s(t)$ is the source signal. When the amplitudes are complex numbers, which is due, for example, to bottom reflections that introduce a phase shift, the propagated signal is~\cite{porter2},
\begin{align}
y(t)=\sum\limits_{k=1}^{K}\Bigl[Re[A_k]s(t-\tau_k)-Im[A_k]s^{+}(t-\tau_k)\Bigr]
\label{eq:signal_prop}
\end{align}
where $s^{+}=\mathcal{H}(s)$ is the Hilbert transform of $s(t)$. The Hilbert transform is a 90 degrees phase shift of $s(t)$ and accounts for the imaginary part of $A_k$. In this case,
\begin{align}
s(t)=&a\cos\Bigl(2\pi\sum\limits_{m=0}^{M-1}f_m t^{m}\Bigr) \\
s^{+}(t)=&a\sin\Bigl(2\pi\sum\limits_{m=0}^{M-1}f_m t^{m}\Bigr) .
\end{align}
\normalsize
\medskip

\section{Detection Models and Statistics}
\label{sec:detection_model}
 It is known that when the noise is additive white Gaussian, the time domain matched filter can achieve optimal detection when the source signal and the propagation environment are known exactly~\cite{johnson1}. When examining the time-frequency performance of a given signal, we can apply the Short Time Fourier Transform to the signal, and generate the spectrogram to analyze it.

\begin{figure}[ht!]
\begin{center}
\includegraphics[height=15mm,width=85mm]{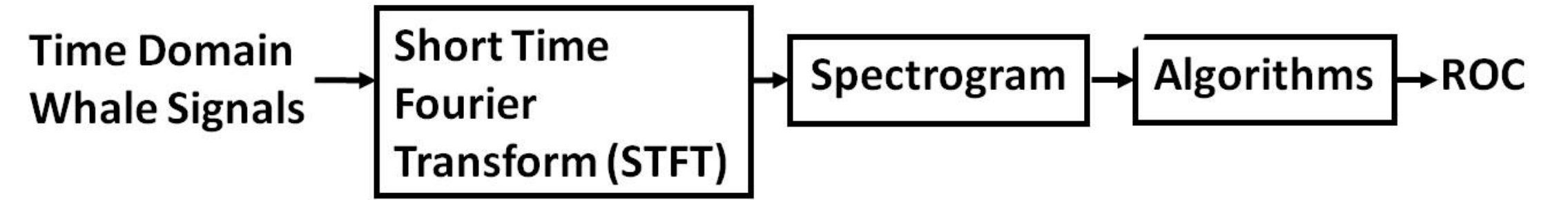}
\end{center}
\vspace{-5pt}
\caption{Detection in time-frequency domain}
\label{fig:procedure}
\end{figure}

 In this section, we will apply the likelihood ratio of the approximated probability distribution of the spectrogram~\cite{atles} and present our STFT detector in the matched ocean, uncertain ocean and mean ocean cases. Simulations of comparison of detection performance can be found in Section~\ref{sec:result}.

\medskip

\subsection{\textbf{Spectrogram Distribution Detector}}
\label{sec:spectrogram_dist}
\subsubsection{Matched ocean}
 We first examine the matched ocean case, that is, the source signal and sound speed profile is known exactly. Our analysis for the matched ocean case is similar to the work of Altes~\cite{atles}.

Let $x(t),~t=0,1,\cdots, N-1$ be the data recorded at the receiver. We want to know whether the data contains the propagated NARW signal. We can form the binary hypothesis test as follows~\cite{StevenKay}
\begin{align}
H_0:~~~x(t)&=n(t), \\
H_1:~~~x(t)&=n(t)+y(t) ,
\end{align}
\normalsize
where $n(t)$ is Gaussian additive noise, $n(t)\sim \mathcal{N}(0,\sigma_n^2)$, and where $y(t)$ is the propagated NARW signal. We apply the STFT to the data and generate the spectrogram.
That is~\cite{flandrin}
\begin{align}
S_x[a,b]&=\biggl|\sum\limits_{i=0}^{M-1}x[aD+i]w[i]\exp(-2\pi jbi/M)\biggr|^2 ,
\end{align}
\normalsize
where $w[i]$ is the window function, $M$ is the length of the window, $D$ is the step of the sliding window, and $a$ is the window shift, $b$ is the frequency shift, and where $\floor{k}$ denotes the greatest integer less than or equal to $k$.

The binary hypothesis test based on the spectrogram is
\begin{align}
H_0:~~~\bold{S_x}&=\bold{S_n}, \\
H_1:~~~\bold{S_x}&=\bold{S_{n+y}} ,
\end{align}
\normalsize
where $\bold{S_x}$ is the spectrogram vector of the data at the receiver, $\bold{S_n}$ is the spectrogram vector of pure noise, and where $\bold{S_{n+y}}$ is the spectrogram vector of the propagated signal plus noise. Let $J=(\floor{(N-M)/D}+1)$, and $B=\floor{M/2}+1$, the spectrogram vector at the receiver is
$\bold{S_x}=[S_x[0,0],\cdots,S_x[J,B]]^T$.
For each spectrogram element $S_x[a,b]$ in $\bold{S_x}$, under the $H_0$ hypothesis, we have~\cite{berger, atles, Martin}
\begin{align*}
p(S_x[a,b])=\frac{1}{M\sigma_n^2}\exp\biggl(-\frac{S_x[a,b]}{M\sigma_n^2}\biggr).
\end{align*}
\normalsize

Under the $H_1$ hypothesis, we have~\cite{atles, Martin, berger}
\begin{align*}
p(S_x[a,b])=&\frac{1}{M\sigma_n^2}\exp\left(-\frac{S_x[a,b]+m_1^2[a,b]+m_2^2[a,b]}{M\sigma_n^2}\right)\\
&\times I_0\left(\frac{2\sqrt{(m_1^2[a,b]+m_2^2[a,b])S_x[a,b]}}{M\sigma_n^2}\right),
\end{align*}
\normalsize
where
\begin{align*}
m_1[a,b]&=\sum\limits_{i=0}^{M-1}y[aD+i]\cos(2\pi bi/M),  \\ m_2[a,b]&=-\sum\limits_{i=0}^{M-1}y[aD+i]\sin(2\pi bi/M),
\end{align*}
\normalsize
and where $I_{\nu}(z)$ is the modified Bessel function of the first kind. Therefore, for each spectrogram element, the likelihood ratio is~\cite{atles, Martin, berger}
\begin{align}
\lambda[a,b]=&\frac{p(S_x[a,b]|H_1)}{p(S_x[a,b]|H_0)} \notag \\
=&\exp\left(-\frac{m_1^2[a,b]+m_2^2[a,b]}{M\sigma_n^2}\right) \nonumber \\
&\times I_0\left(\frac{2\sqrt{(m_1^2[a,b]+m_2^2[a,b])S_x[a,b]}}{M\sigma_n^2}\right) . \label{eq:lambda_chi_element}
\end{align}
\normalsize

Assuming that each element in the spectrogram matrix is statistically independent, the likelihood ratio is
\begin{align}
\lambda=&\frac{p(\bold{S_x}|H_1)}{p(\bold{S_x}|H_0)}=\frac{p(S_x[0,0]|H_1)\cdots p(S_x[J,B]|H_1)}{p(S_x[0,0]|H_0)\cdots p(S_x[J,B]|H_1)} \notag \\
=&\prod\limits_{a=0}^{J}\prod\limits_{b=0}^{B}\frac{p(S_x[a,b]|H_1)}{p(S_x[a,b]|H_0)}=\prod\limits_{a=0}^{J}\prod\limits_{b=0}^{B}\lambda[a,b] . \label{eq:lambda_approx1}
\end{align}
\normalsize

\medskip

\subsubsection{Uncertain ocean}
In reality, we may not know the environment and sound speed profile of the ocean exactly. If we know the prior distribution of the sound speed profile, we can apply the Bayes rule, and incorporate the prior information into detection. Suppose the sound speed profile is uniformly distributed over $P$ possible cases, the likelihood ratio, based on eq.~(\ref{eq:lambda_approx1}), in this case is
\begin{align}
\lambda_u=&\frac{1}{P}\sum\limits_{k=1}^{P}\Bigl(\prod\limits_{a=0}^{J}\prod\limits_{b=0}^{B}\lambda_k[a,b]\Bigr) \notag \\
=&\frac{1}{P}\sum\limits_{k=1}^{P}\Biggl(\prod\limits_{a=0}^{J}\prod\limits_{b=0}^{B}\biggl(\exp\Bigl(-\frac{m_{k1}^2[a,b]+m_{k2}^2[a,b]}{M\sigma_n^2}\Bigr) \notag \\
&\times I_0\Bigl(\frac{2\sqrt{(m_{k1}^2[a,b]+m_{k2}^2[a,b])S_{kx}[a,b]}}{M\sigma_n^2}\Bigr)\biggr)\Biggr) , \label{eq:lambda_approx_uncertain}
\end{align}
\normalsize
where $\lambda_k$ is the likelihood ratio of the spectrogram of the $k^{th}$ possible propagated signal, $S_{kx}[a,b]$ is the spectrogram element of the $k^{th}$ possible propagated signal, and where $m_{k1}[a,b]$ and $m_{k2}[a,b]$ are the mean of real and imaginary part respectively of the $k^{th}$ possible propagated signal.

\medskip

\subsubsection{Mean ocean}
When the sound speed profile is uncertain, and we only have the knowledge of the mean value of the possible sound speed profiles, which is mismatched with the true sound speed profile. Based on eq.~(\ref{eq:lambda_approx1}), the likelihood ratio in this case becomes~\cite{book1997}
\begin{align}
\lambda_m=&\prod\limits_{a=0}^{J}\prod\limits_{b=0}^{B}\Biggl(\exp\biggl(-\frac{\hat{m}_1^2[a,b]+\hat{m}_2^2[a,b]}{M\sigma_n^2}\biggr) \nonumber \\ &\times I_0\biggl(\frac{2\sqrt{(\hat{m}_1^2[a,b]+\hat{m}_2^2[a,b])S_x[a,b]}}{M\sigma_n^2}\biggr)\Biggr) , \label{eq:lambda_chi_element3}
\end{align}
where $S_x[a,b]$ is the spectrogram element of the received signal, and $\hat{m}_1[a,b]$ and $\hat{m}_2[a,b]$ are the mean value of the real part and imaginary part respectively of the Short Time Fourier Transform of the signal in the mean ocean environment, and where the sound speed profile is assumed to take average values.

\medskip

\subsection{\textbf{STFT Detector}}
\label{sec:stft_detector}
The detection algorithms based on spectrogram data neglect phase information, and therefore their detection performances are not likely to be optimal. In this section we derive the probability distribution and likelihood ratio for detection of STFT. The STFT detector preserves the phase information of the detected signal, and achieves optimal detection performance in the matched ocean case.

\medskip

\subsubsection{Matched ocean}
The binary hypothesis test based on the STFT in this case is~\cite{StevenKay}
\begin{align}
H_0:~~~X&=X_n, \\
H_1:~~~X&=X_y+X_n,
\end{align}
where $X$ is the vectorized STFT matrix of the received data, $X_n$ is the vectorized STFT matrix of pure noise, and $X_y$ is the vectorized STFT matrix of the propagated signal. $X=[X[0,0], \cdots,X[J,B]]^{T}$. For each STFT element $X[a,b]$ in $X$, we have~\cite{flandrin}
\begin{align}
X[a,b]=\sum\limits_{i=0}^{M-1}x[aD+i]w[i]\exp(-2\pi jbi/M) ,
\label{eq:stft}
\end{align}
 where $a=0,\cdots, J$ and $b=0,\cdots, B$. Since $X[a,b]$ is a complex value, letting
\begin{align}
U[a,b]&=Re\{X[a,b]\} \label{eq:stft_real} , \\
V[a,b]&=Im\{X[a,b]\} \label{eq:stft_imag} ,
\end{align}
we have $X[a,b]=U[a,b]+jV[a,b]$.

Concatenating the elements in the STFT matrix of the received data, and separating the real and imaginary part of each element, we can obtain
\begin{align}
\bold{X}=(U[0,0],\cdots,U[J,B],V[0,0],\cdots,V[J,B])^T . \label{eq:x_stft}
\end{align}

 Because we use the Gaussian distribution as our noise model, $\bold{X}$ follows a multivariate normal distribution under both $H_0$ and $H_1$ hypotheses. It can be proved that the signal covariance matrix $\bold{C_x}$ is the same under both $H_0$ and $H_1$ hypotheses. The expression for $\bold{C_x}$ is derived in Appendix~\ref{sec:deriv_stft}.

When $\bold{C_x}$ is singular, the probability density function for $\bold{X}$ does not exist. According to the derivation in Appendix~\ref{sec:deriv_stft}, mapping $\bold{X}$ into a subspace formed by $Q_1$, where $Q_1$ is the eigenvector of the non-zero singular values' component, we can obtain the probability density function for $Q_1^TX$ under $H_0$ and $H_1$ hypotheses
\begin{align}
p(Q_1^T\bold{X}|H_0)=&\frac{1}{(2\pi)^{(B+1) (J+1)}\det^{1/2}(\Lambda_1)} \nonumber \\ &\times \exp\biggl(-\frac{1}{2}\bold{X^T}Q_1\Lambda_1^{-1}Q_1^T\bold{X}\biggr) \label{eq:stft1_h0} , \\
p(Q_1^T\bold{X}|H_1)=&\frac{1}{(2\pi)^{(B+1) (J+1)}\det^{1/2}(\Lambda_1)} \nonumber \\
&\times \exp\biggl(-\frac{1}{2}\bold{(X-\mu)^T} Q_1\Lambda_1^{-1}Q_1^T\bold{(X-\mu)}\biggr). \label{eq:stft1_h1}
\end{align}
\normalsize
where $\Lambda_1$ is the diagonal matrix of eigenvalues corresponding to $Q_1$, and $\mu$ is the expected value of $\bold{X}$.

The likelihood ratio based on the STFT is then
\begin{align}
\lambda=&\frac{p(Q_1^T\bold{X}|H_1)}{p(Q_1^T\bold{X}|H_0)} \notag \\
=&\exp\Bigl(-\frac{1}{2}(\bold{\mu^T} Q_1\Lambda_1^{-1}Q_1^T\bold{\mu}-2\bold{X^T}Q_1\Lambda_1^{-1}Q_1^T\bold{\mu})
\Bigr).
\label{eq:lambda_stft}
\end{align}

 We can calculate the analytic solution for the ROC plot based on eq.~(\ref{eq:lambda_stft}). According to the derivation in Appendix~\ref{sec:deriv_stft}, the square of separation parameter, or the detection index is
\begin{align}
d^2&=\bold{\mu^T}Q_1\Lambda_1^{-1}Q_1^T\bold{\mu}. \label{eq:stft_ana_ske}
\end{align}

\medskip

\subsubsection{Uncertain ocean}
When the sound speed profile is uncertain, and is uniformly distributed over $P$ possible cases. Applying the Bays rule, the likelihood ratio can be calculated as
\begin{align}
\lambda_u=&\frac{1}{P}\sum\limits_{k=1}^{P}\exp\Bigl(\bold{X^T}Q_1\Lambda_1^{-1}Q_1^T\bold{\mu_{k}}-\frac{1}{2}\bold{\mu_{k}^T}Q_1\Lambda_1^{-1}Q_1^{T}\bold{\mu_{k}}\Bigr) ,\label{eq:stft_uncertain}
\end{align}
\normalsize
where $\bold{\mu_k}$ is the expected value of the $k^{th}$ possible propagated signal's STFT vector.

\medskip

\subsubsection{Mean ocean}
When the sound speed profile is uncertain, and the prior information of the sound speed profile is unknown, but we know the mean sound speed profile, we can cross-correlate the possible propagated signals with the signal corresponding to the mean sound speed profile environment. The likelihood ratio can be computed as~\cite{book1997}
\begin{align}
\lambda_m=&\exp\biggl(-\frac{1}{2}\Bigl(\bold{\mu_m^T} Q_1\Lambda_1^{-1}Q_1^T\bold{\mu_m}-2\bold{X^T}Q_1\Lambda_1^{-1}Q_1^T\bold{\mu_m}\Bigr)
\biggr),
\label{eq:lambda_stft_mean}
\end{align}
where $\bold{\mu_m}$ is the expected value of the mean sound speed profile with respect to $\bold{X}$. We can mathematically prove that when the uncertainty of the sound speed profile is small, given that the mapping of sound speed profile to the propagated signal is linear, according to eq.~(\ref{eq:signal_prop}), eq.~(\ref{eq:lambda_stft_mean}) is actually the geometric mean of the likelihood ratio. Detailed proof can be found in Appendix~\ref{sec:deriv_stft}.

\medskip

\section{Results}
\label{sec:result}
\begin{figure}[ht!]
     \begin{center}
        \subfigure[Waveform]{%
            \includegraphics[width=.235\textwidth, height=3.6cm]{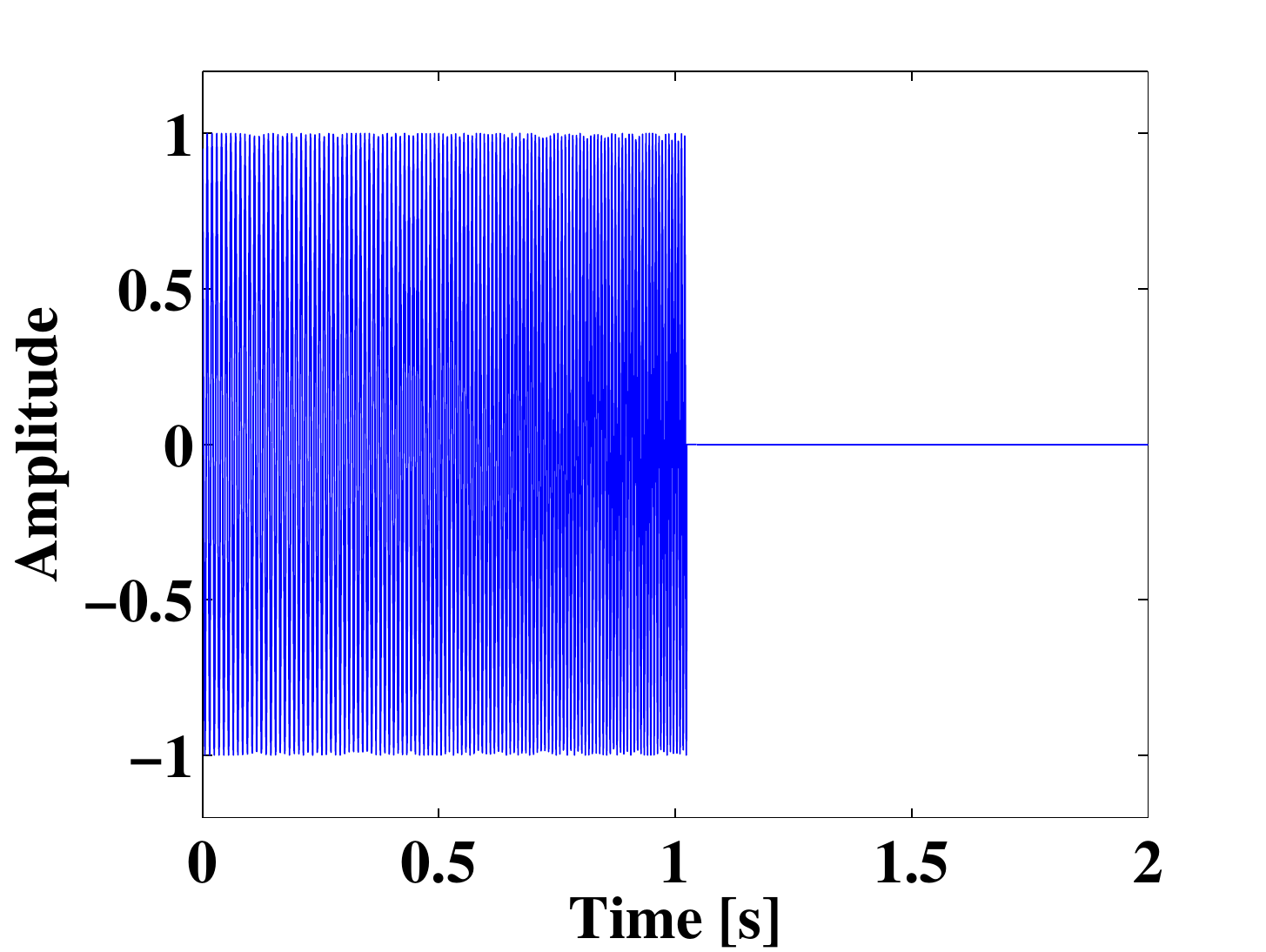}
        }%
        \subfigure[Spectrogram]{%
           \includegraphics[width=.235\textwidth, height=3.6cm]{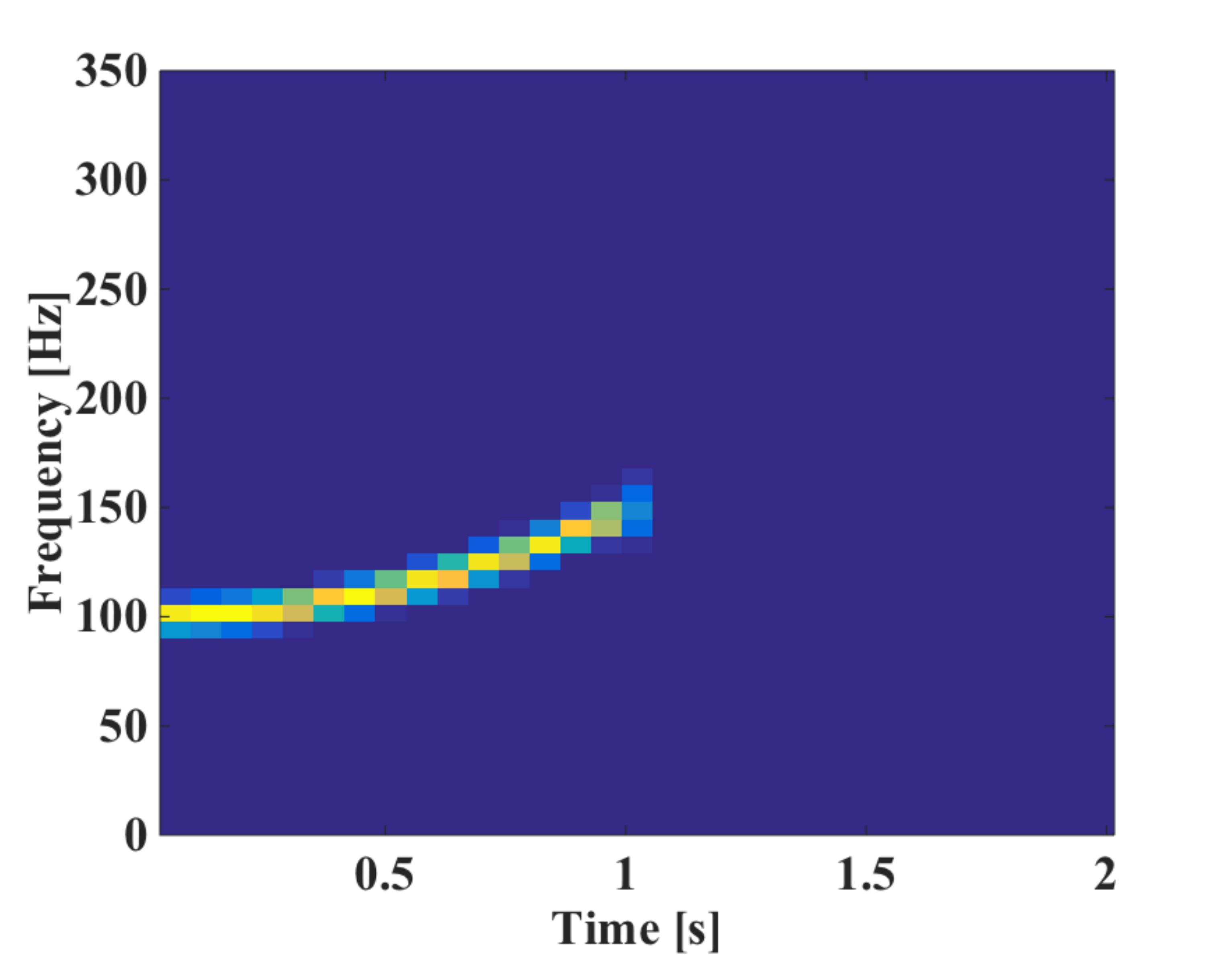}
        }
    \end{center}
    \vspace{-10pt}
    \caption{Synthetic NARW source signal}\label{fig:sig}
\end{figure}

Based on recordings of NARW from the years 2001 to 2003 in the Cape Cod Bay region, using bottom-mounted hydrophones with 2000Hz sampling rate, the polynomial coefficient set $(f_0, f_1, f_2)$ was found most frequently to take the value of (100,0,48)~\cite{clark1}. Because the NARW upsweep call usually lasts for about 1 second, we let the duration of the signal be 1.024s. Substituting these values into eq.~(\ref{eq:signal_model}), and letting $a=1$, the source signal's waveform and spectrogram can be determined and are illustrated in Figure~\ref{fig:sig}.

We place the source signal 15 meters below the sea surface, and place the receiver 35 meters below. The horizontal distance between the receiver and source signal is 1000 meters. Considering the uncertainty of the sound speed profile, we suppose that the sound speed profile has 50 possible cases. When the sound speed profile changes, the propagated signal will also change. Figure 6 illustrates an example that a source under two distinct sound speed profiles will produce two different propagated signals.
\begin{figure}[ht!]
\centering
\subfigure{
\includegraphics[height=34mm,width=41mm]{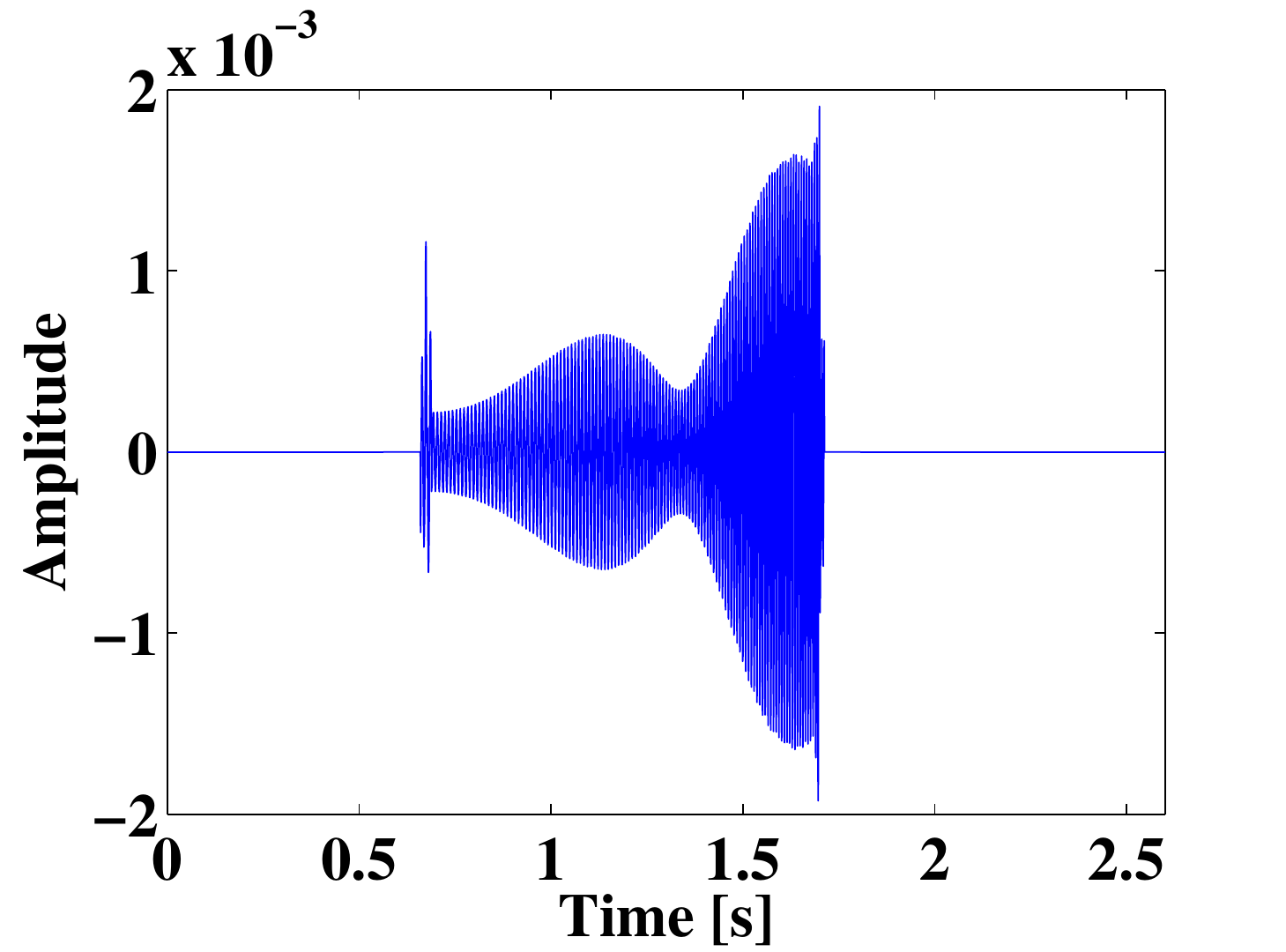}}
\subfigure{
\includegraphics[height=34mm,width=41mm]{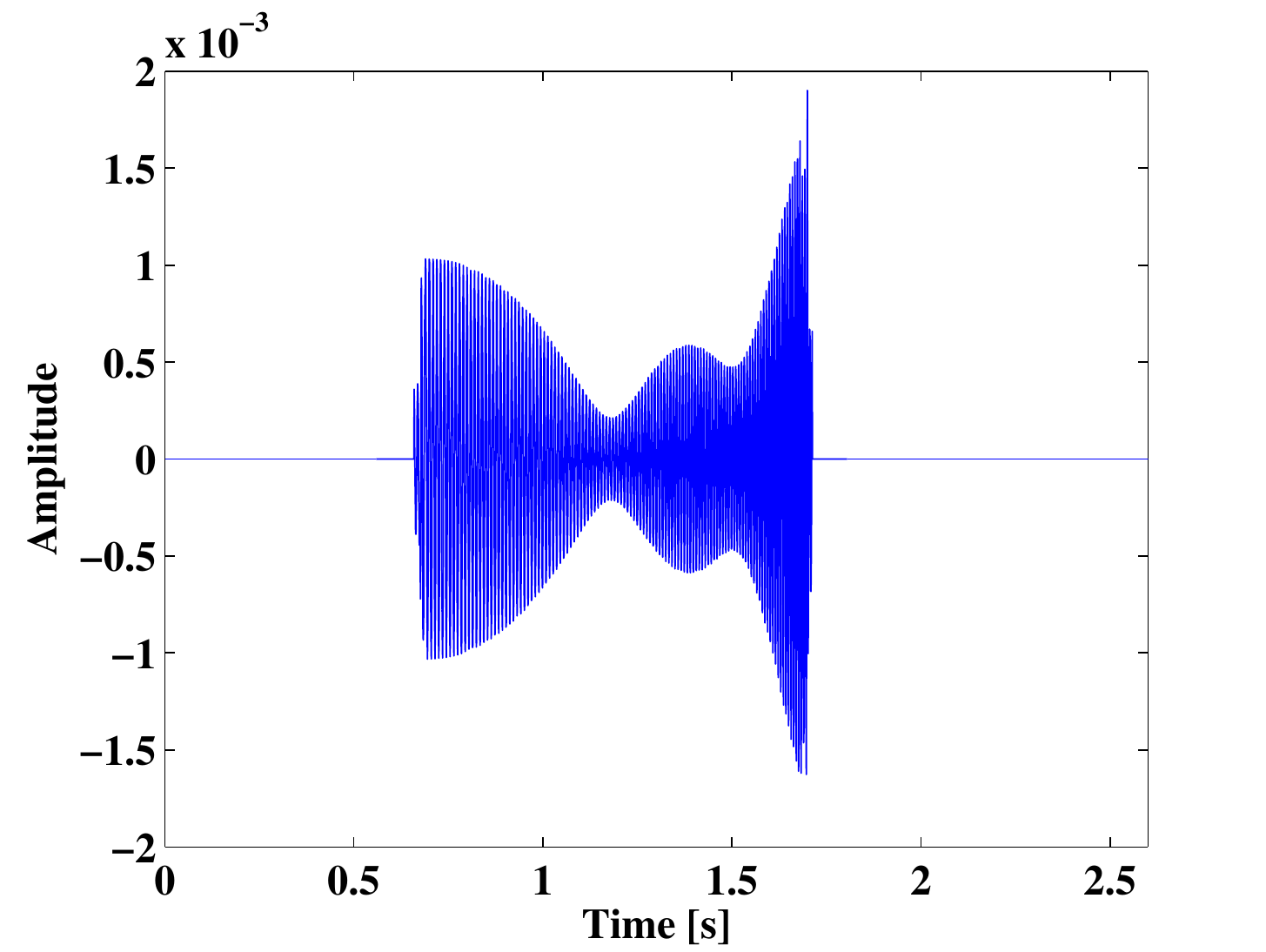}
}
\\
\subfigure{
\includegraphics[height=34mm,width=41mm]{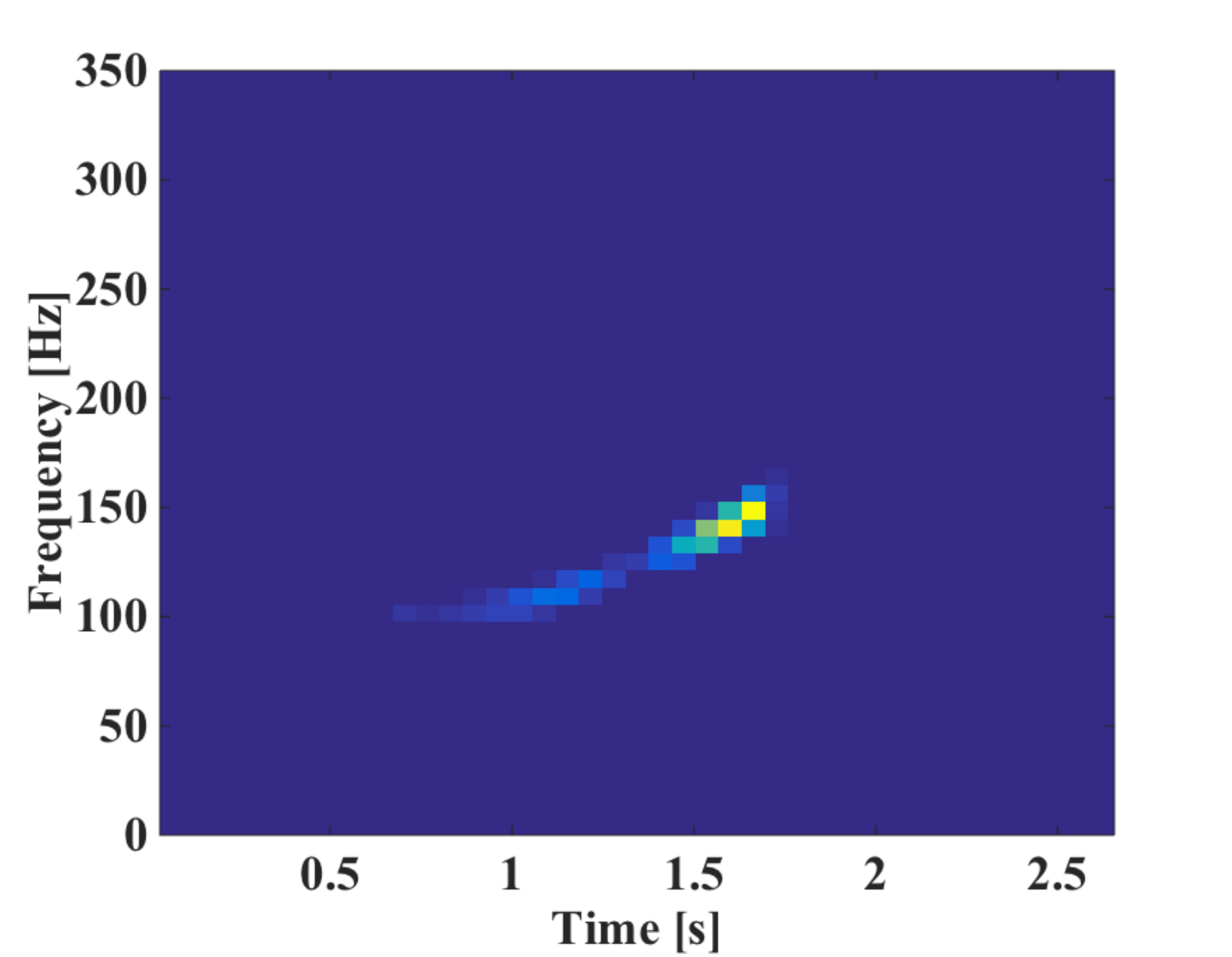}
}
\subfigure{
\includegraphics[height=34mm,width=41mm]{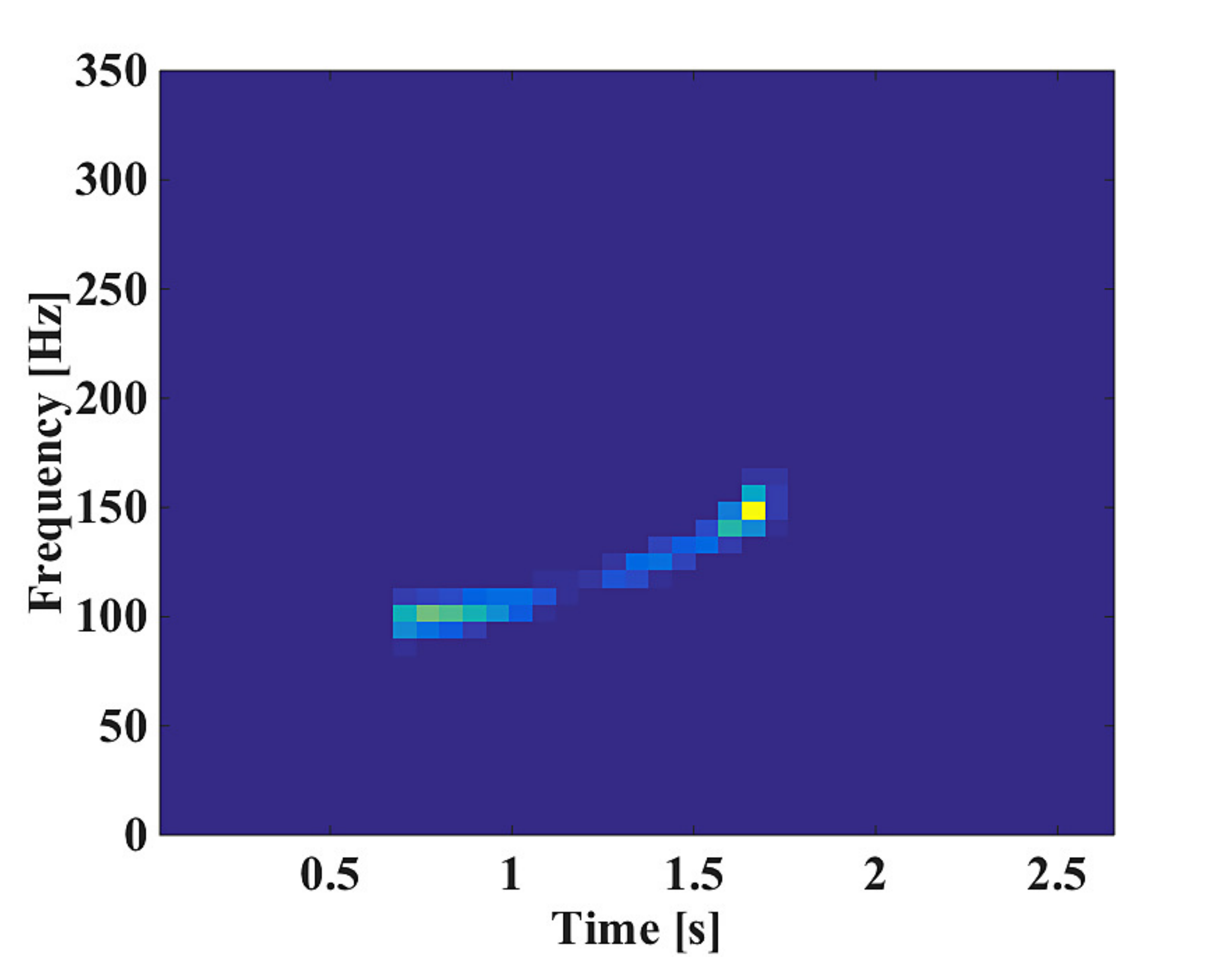}
}
\vspace{-5pt}
\caption{Example plots of time domain waveform and spectrogram of NARW propagated signal under two different sound speed profiles.}
\label{fig:narw_prop}
\end{figure}

According to the spectrogram plots, we can see that there is a dispersion effect during signal propagation. Some frequency components in the source signal have been lost due to phase cancellation effects of multipath propagation environment. However, the overall shapes of 50 possible propagated signal spectrograms are similar to the source signal spectrogram; this is because the time shift will not change the frequency content of the signal, due to the basic property of the Fourier Transform.

\medskip

\subsection{\textbf{Detectors Tested}}
We compare the detection performance of the STFT detector, the spectrogram distribution detector, and the spectrogram correlation detector in the matched ocean, uncertain ocean and mean ocean cases. We benchmark the results with the optimal detection performance given by the time domain matched filtering assuming that the source signal and propagation environment are known exactly.

The spectrogram correlation detector is a well known work in marine mammal acoustic detection field. Review of the spectrogram correlation detector~\cite{mellinger1} proposed by Mellinger and Clark is as follows. The spectrogram correlation constructs a kernel function for the vocalization signal, then cross correlates it with the target signal's spectrogram to calculate the recognition score, and does detection based on the recognition score. The kernel function for the signal is made up of several segments, one per FM section in the target vocalization type. The kernel value $k$ at a given time and frequency point $(t,f)$ is specified by:
\begin{align*}
x&=f-\biggl(f_0+\frac{t}{d}(f_1-f_0)\biggr) \\
k(t,f)&=\biggl(1-\frac{x^2}{\sigma^2}\biggr)\exp\biggl(-\frac{x^2}{2\sigma^2}\biggr)
\end{align*}
\normalsize
where $x$ is the distance of the point $(t,f)$ from the central axis of the segment at time $t$, $f_0$ is the start frequency of the segment, $f_1$ is the end frequency of the segment, $d$ is the duration of the segment, and $\sigma$ is the instantaneous bandwidth of the segment at time $t$. The recognition score is calculated by cross correlating the kernel $k(t,f)$ with the spectrogram of the signal:
\begin{align*}
\alpha(t)=\sum\limits_{t_0}\sum\limits_{f}k(t_0,f)S(t-t_0,f)
\end{align*}

We use a rectangular window function with length 256, and overlap size 128 to generate the spectrogram of the signal. We apply these parameters to the spectrogram correlation detector and spectrogram distribution detector. We set a threshold for the recognition score and generate the Receiver Operating Characteristic (ROC) curve.

\begin{figure*}
    \begin{center}
     \subfigure[Matched ocean, SNR=4]{%
            \includegraphics[width=.25\textwidth, height=3.6cm]{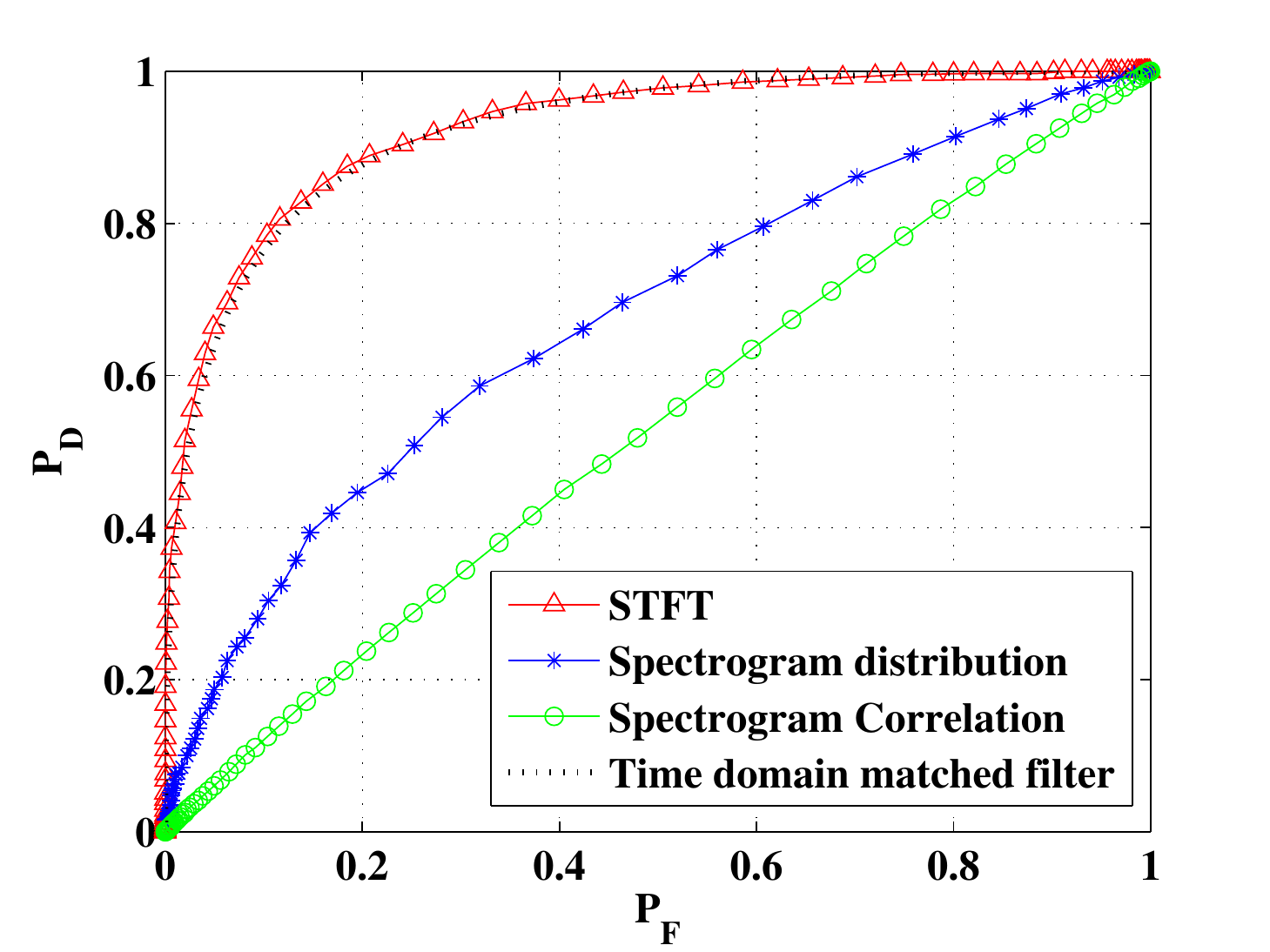}
        }%
         \subfigure[Uncertain ocean, SNR=4]{%
            \includegraphics[width=.25\textwidth, height=3.6cm]{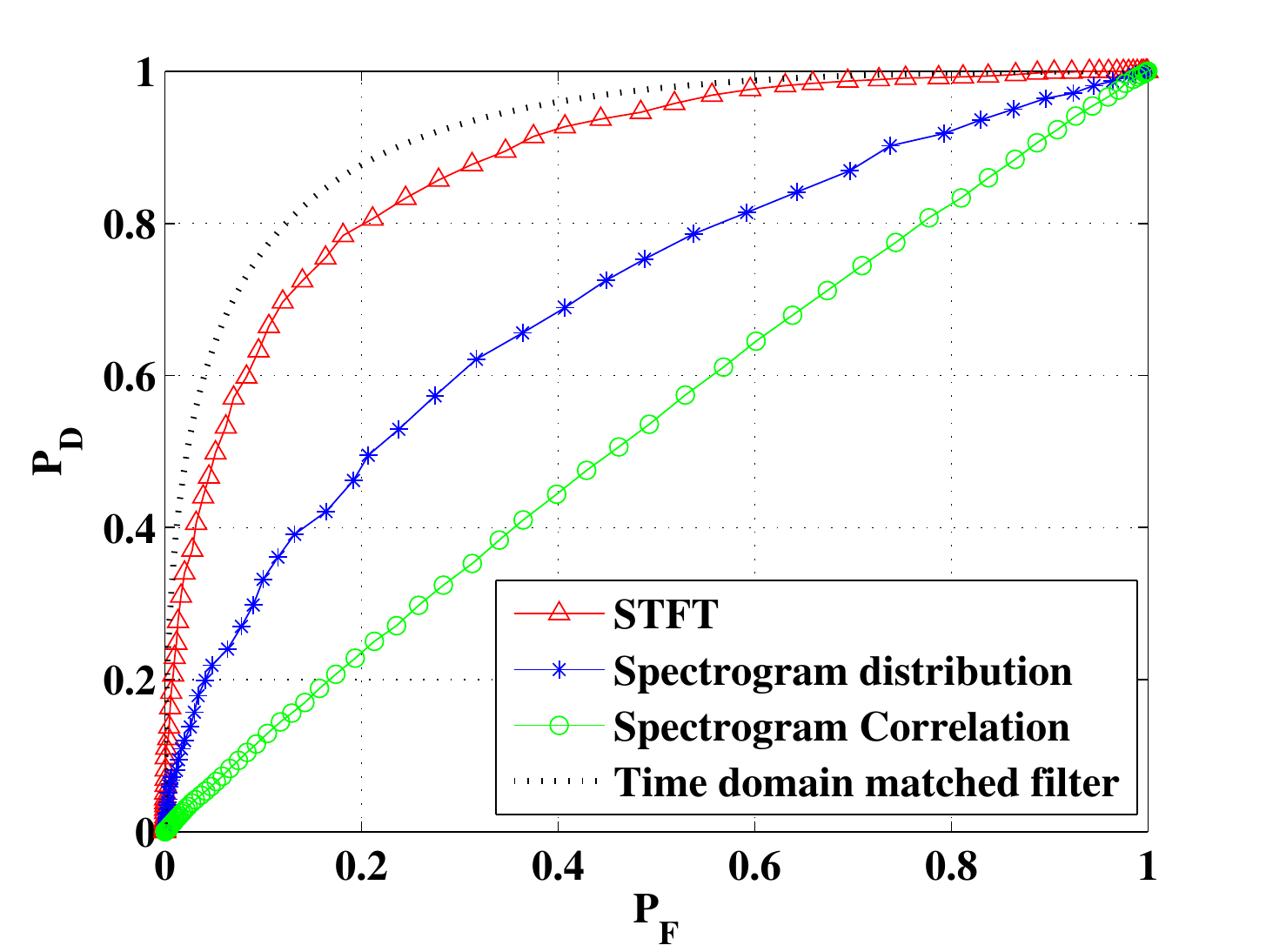}
        }
        \subfigure[Mean ocean, SNR=4]{%
           \includegraphics[width=.25\textwidth, height=3.6cm]{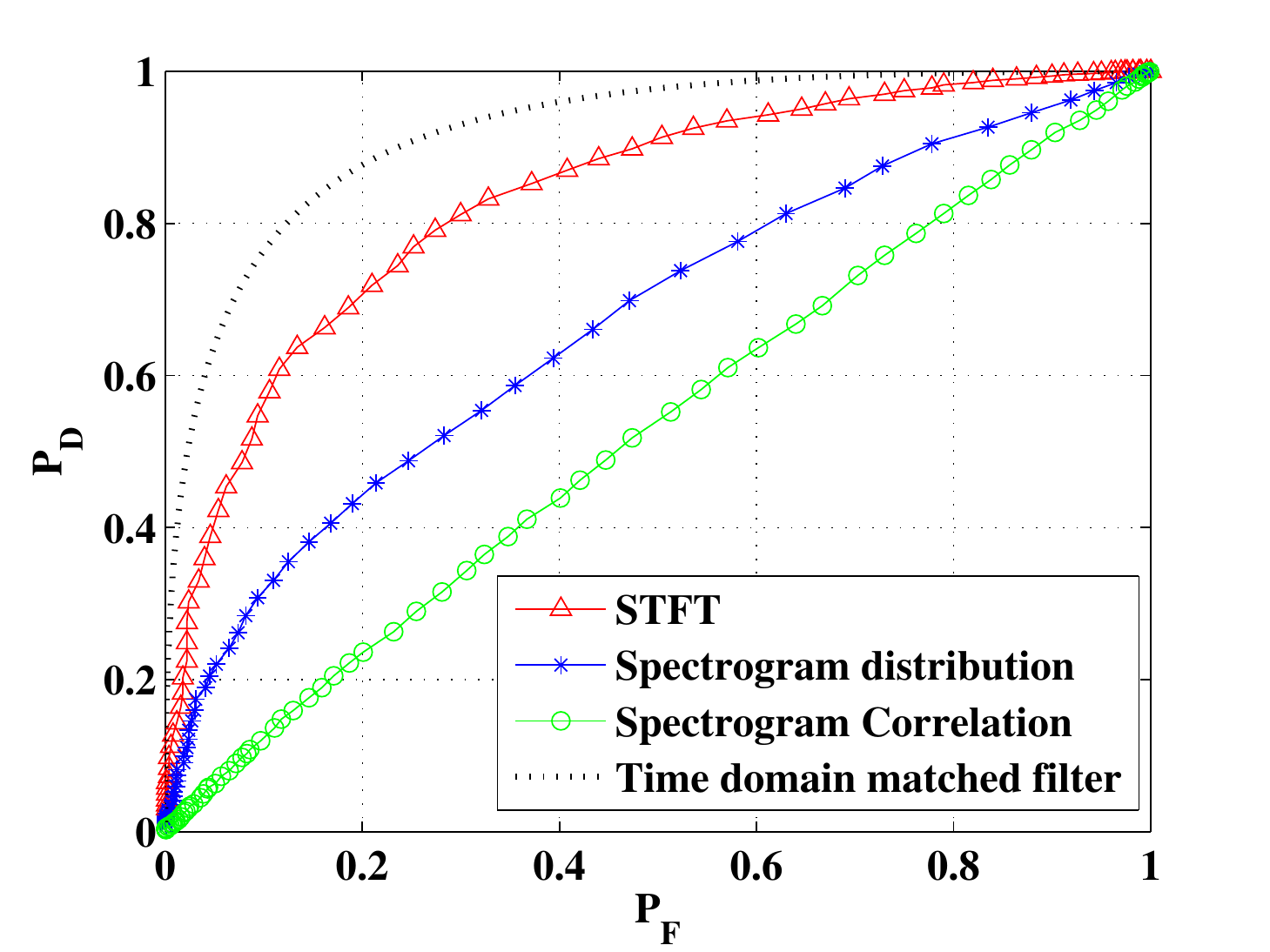}
        }\\
        \subfigure[Matched ocean, SNR=16]{%
           \includegraphics[width=.25\textwidth, height=3.6cm]{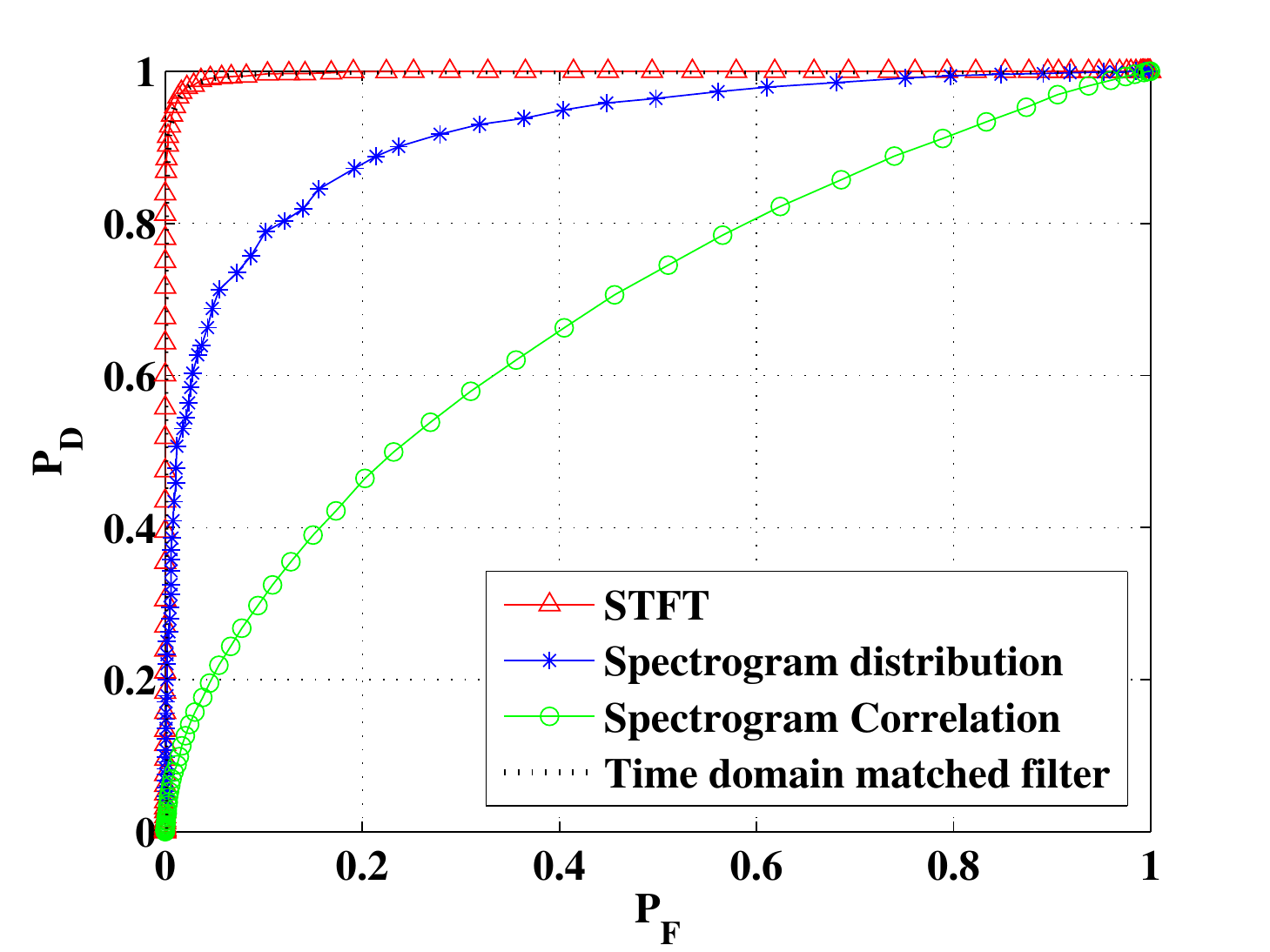}
        }
        \subfigure[Uncertain ocean, SNR=16]{%
            \includegraphics[width=.25\textwidth, height=3.6cm]{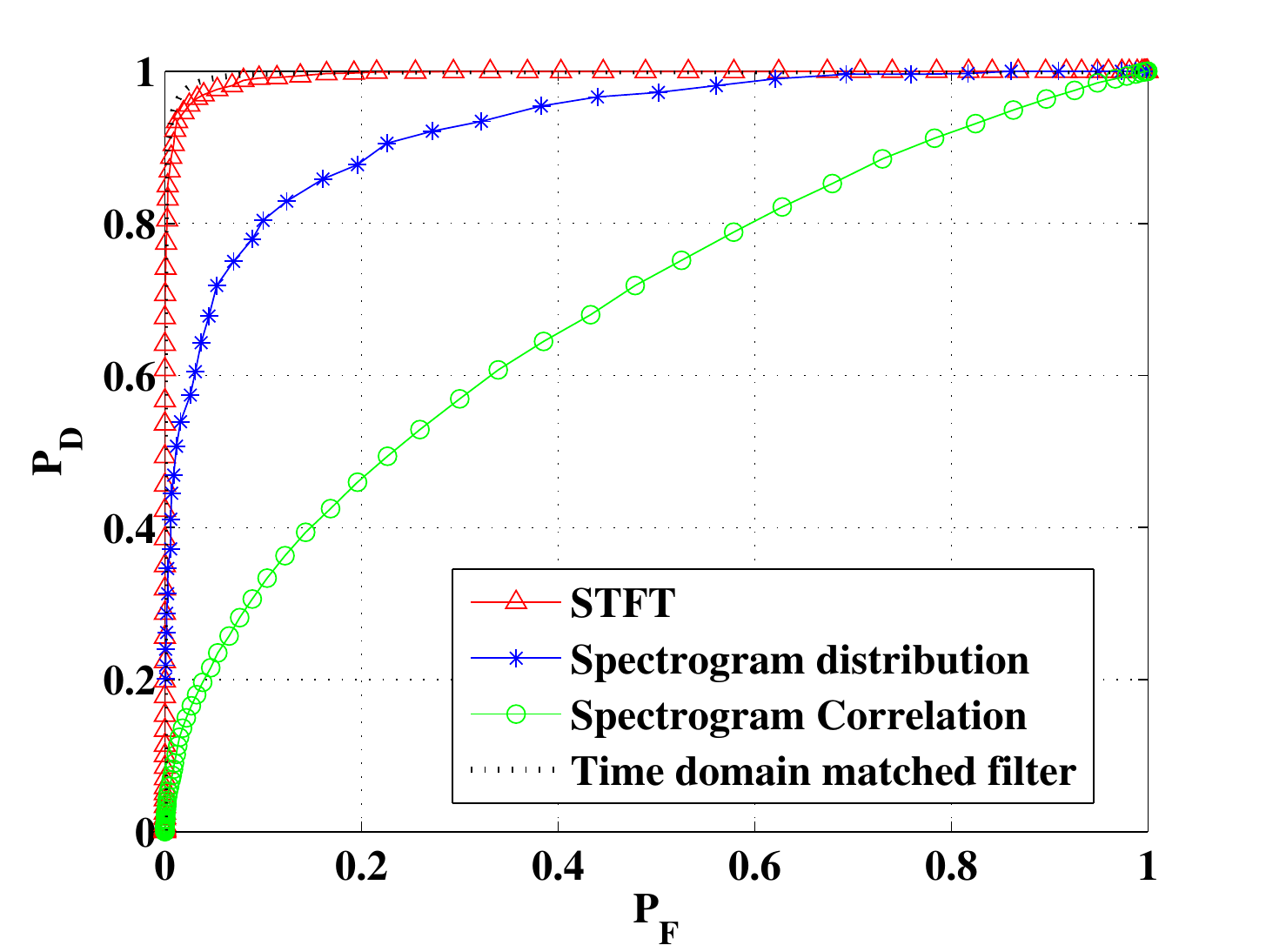}
        }%
        \subfigure[Mean ocean, SNR=16]{%
            \includegraphics[width=.25\textwidth, height=3.6cm]{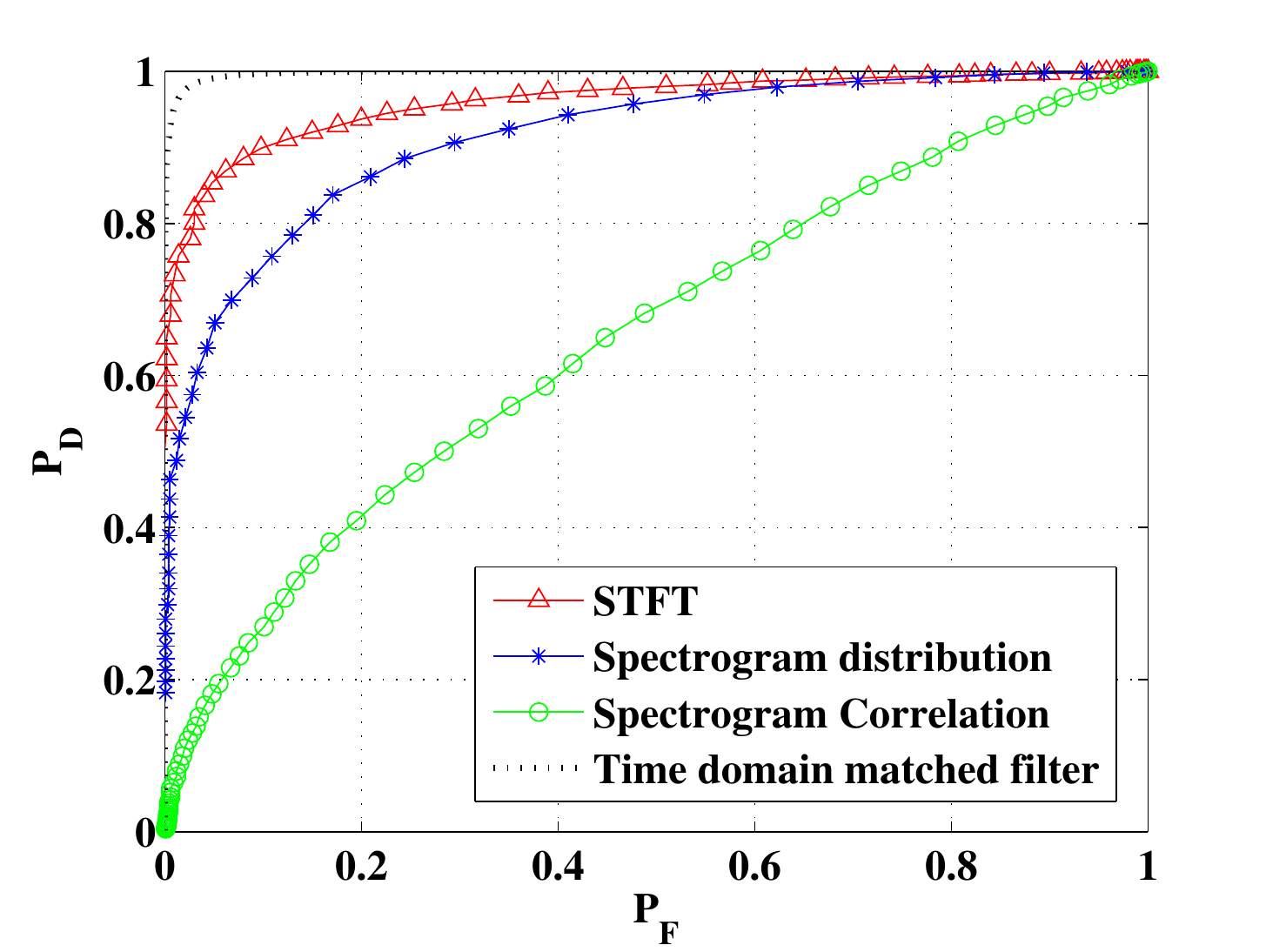}
        }%
    \end{center}
    \caption{
        ROC plots for the detectors in various ocean environments and SNRs. (a), (b) and (c) show the detectors performances of the matched ocean, uncertain ocean and mean ocean environment case when SNR=4; (d), (e) and (f) show the matched ocean, uncertain ocean and mean ocean environment case when SNR=16.
     }%
   \label{fig:roc_detectors}
\end{figure*}

The comparison of detection performance of the STFT detector, the spectrogram distribution detector and the spectrogram correlation detector is shown in Figure~\ref{fig:roc_detectors}. We use a Monte Carlo method to do the numerical simulation. We will analyze the detection performance in the matched ocean, uncertain ocean and mean ocean cases.

\medskip

\subsection{\textbf{Matched Ocean}}
\begin{figure}[!ht]
\begin{center}
\includegraphics[height=53mm,width=80mm]{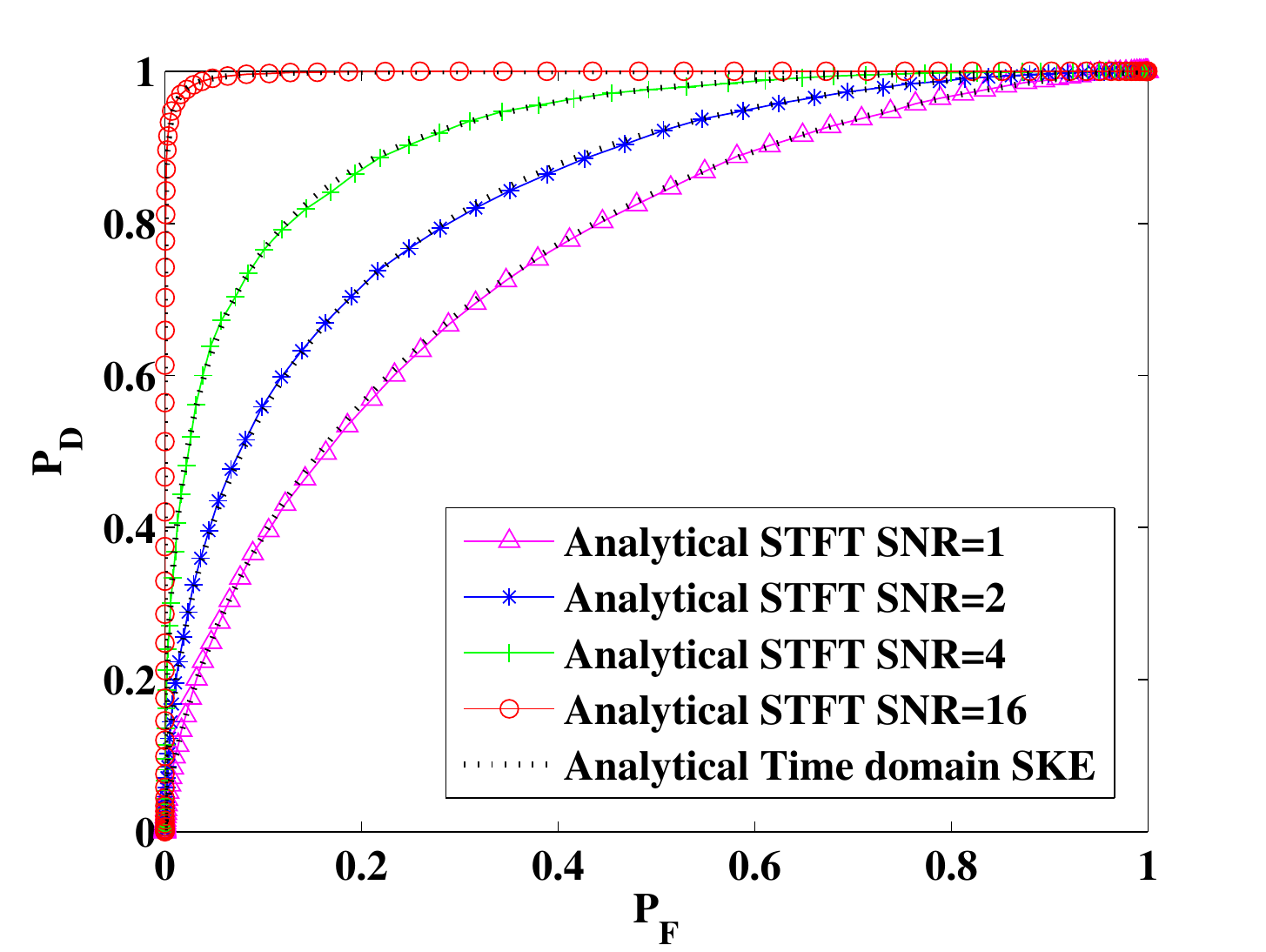}
\end{center}
\caption{ROC plot based on analytical derivation when the environmental parameters are known exactly in different SNRs}
  \label{fig:stft_ske}
\end{figure}

We first examine the special case of detection in a matched ocean propagation environment. We can see that in this case the STFT detector achieves identical detection performance to the time domain matched filter case, which is the optimal detection performance. We used different window lengths and amounts of overlap, and the STFT detector performs identically in all situations. Figure~\ref{fig:stft_ske} shows the analytic ROC plots in different SNRs for the STFT detector, based on the separation index in eq.~(\ref{eq:stft_ana_ske}) and the analysis in Appendix~\ref{sec:deriv_stft}; along with the corresponding analytic time domain matched filtered ROC curves as a comparison. We can see the analytic derivation and the numerical simulation are a good match.

Examining Figure~\ref{fig:roc_detectors}, the STFT detector performs better than both the spectrogram distribution detector and the spectrogram correlation detector under different SNRs since it preserves the phase information. The spectrogram distribution detector performs better than the spectrogram correlation detector, and the spectrogram correlation detector does not perform well especially at low SNR. This is because the multipath effect will lead to dispersion for the signal, and some energy of the signal will be lost due to phase cancelation. When such a propagated signal is corrupted by noise, the performance of the spectrogram correlation detector suffers further, because it doesn't exploit the probability distribution of the spectrogram elements and noise.

\medskip

\subsection{\textbf{ Uncertain Ocean}}
\begin{figure}
\begin{center}
\includegraphics[height=53mm,width=80mm]{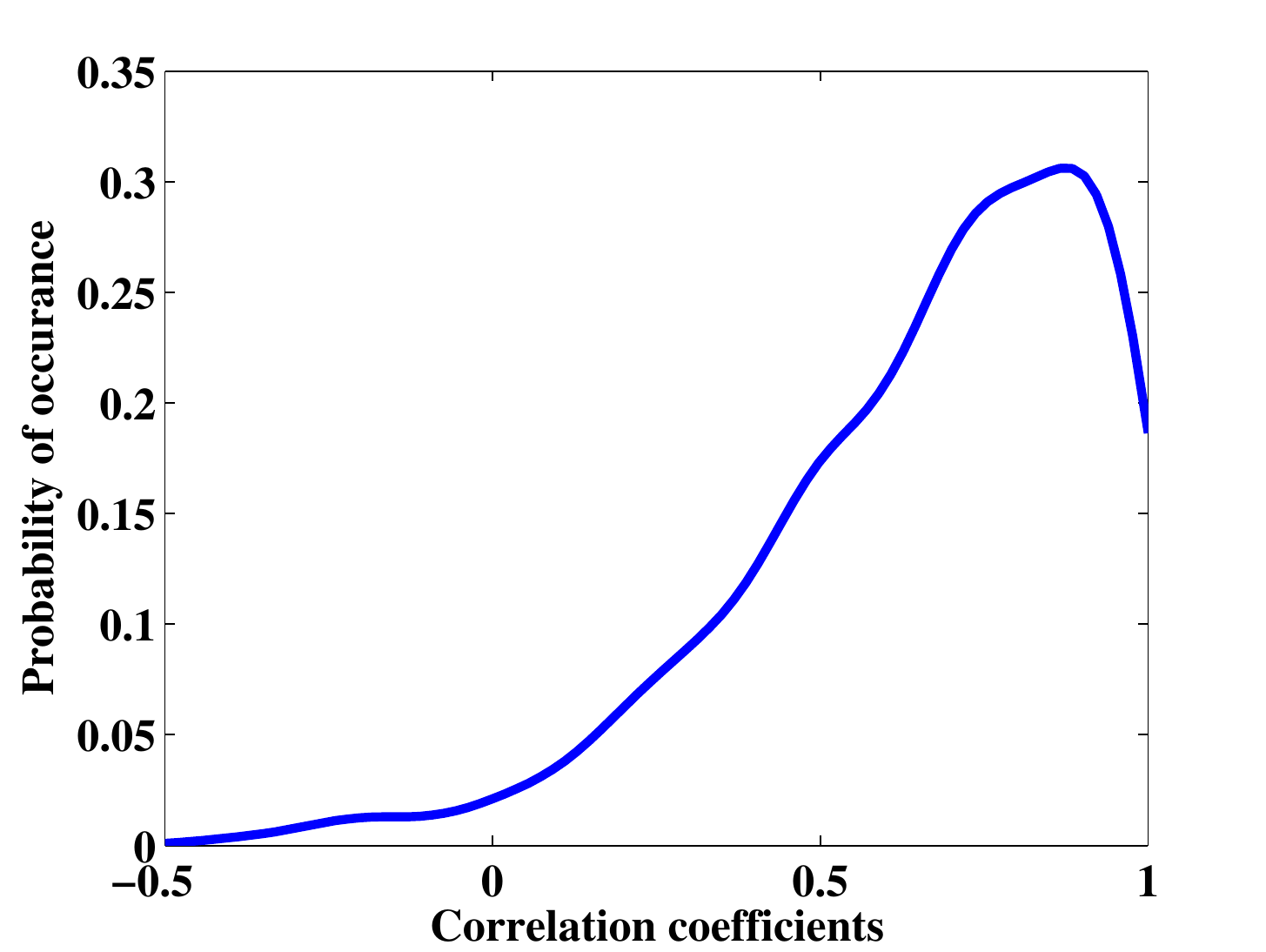}
\end{center}
\caption{Kernel density function of propagated signals' correlation coefficients of different sound speed profiles}
\label{fig:hist_coeff}
\end{figure}

For the uncertain ocean case, we assume there are 50 possible sets of sound speed profiles. The values of the sound speed profile are given in Table I. From the ROC plots, we can see that the STFT detector performs close to the optimal case given by the time domain matched filter; and its detection performance is much better than the spectrogram distribution detector and spectrogram correlation detector. The kernel density function of correlation coefficients of the propagated signals of each possible sound speed profile is shown in Figure~\ref{fig:hist_coeff}. We can see that most of the correlation coefficients are above 0.5; therefore, when we assume a uniform distribution over the possible sound speed profiles, the STFT detector performs close to the optimum.

\medskip

\subsection{\textbf{Mean Ocean}}
For the mean ocean case, we can see that the overall detection performance deteriorates. Comparing the figures for the matched ocean case, the uncertain ocean case and the mean ocean case, we can see that the detection performance for the matched ocean case is the best, and the uncertain ocean is better than the mean ocean case. This is because the matched ocean has the most information while the mean ocean has the least. The performance of the spectrogram distribution detector does not change as much as the STFT detector, and there is little change in the performance of the spectrogram correlation detector for the three different environment cases. This implies that, by neglecting phase information, the detector will be less sensitive to environmental uncertainty.

\medskip

\section{Conclusion}
\label{sec:conclusion}
In this chapter, we have presented the STFT detector, which preserves the phase information of the signal. We have also determined the likelihood ratio of the STFT detector and the spectrogram distribution detector in the matched ocean, uncertain ocean and mean ocean cases. Experiments show that the STFT detector performs better than the spectrogram distribution detector and spectrogram correlation detector. The STFT detector is more sensitive to environmental changes, because it includes the phase information, while the spectrogram based detectors are less sensitive. By exploiting the probability distribution of noise and spectrogram element, the detector can be improved and be more robust to multipath propagation environment.

\medskip

\appendices
\section{Derivation of Short time Fourier Transform Detector}
\label{sec:deriv_stft}
\subsection{\textbf{Matched ocean}}
According to eq.~(\ref{eq:x_stft}), we can see that $\bold{X}$ is of dimension $(2(B+1)\times (J+1))\times 1$. Because we use the Gaussian distribution as our noise model, $\bold{X}$ follows a multivariate normal distribution under both $H_0$ and $H_1$ hypotheses.

Under the $H_0$ hypothesis, we have
\small
\begin{align}
\mathbb{E}(\bold{X})=\bold{0}_{2(B+1)\times (J+1))\times 1} .
\end{align}
\normalsize

The covariance matrix $\bold{C_x}$ of $\bold{X}$ can be represented as~\cite{StevenKay2}
\small
\begin{align}
\bold{C_x}=\left[
\begin{array}{cc}
\bold{C_{UU}} &\bold{C_{UV}} \label{eq:cov_mat} \\
\bold{C_{VU}} &\bold{C_{VV}}
\end{array}
\right] ,
\end{align}
\normalsize
where
\small
\begin{align*}
\bold{C_{UU}}=\left[
\begin{array}{ccc}
Var(U[0,0]) & \cdots & Cov(U[0,0],U[J,B]) \\
\vdots &\ddots & \vdots \\
Cov(U[J,B],U[0,0]) &\cdots & Var(U[J,B])
\end{array}
\right] ,
\end{align*}
\begin{align*}
\bold{C_{VV}}=\left[
\begin{array}{ccc}
Var(V[0,0]) & \cdots & Cov(V[0,0],V[J,B]) \\
\vdots &\ddots & \vdots \\
Cov(V[J,B],V[0,0]) &\cdots & Var(V[J,B])
\end{array}
\right] ,
\end{align*}
\begin{align*}
\bold{C_{UV}}=\left[
\begin{array}{ccc}
Cov(U[0,0],V[0,0]) &\cdots & Cov(U[0,0],V[J,B]) \\
\vdots &\ddots & \vdots\\
Cov(U[J,B],V[0,0]) &\cdots & Cov(U[J,B],V[J,B])
\end{array}
\right] ,
\end{align*}
\begin{align*}
\bold{C_{VU}}=\left[
\begin{array}{ccc}
Cov(V[0,0],U[0,0])  &\cdots & Cov(V[0,0],U[J,B]) \\
\vdots &\ddots & \vdots \\
Cov(V[J,B],U[0,0])  &\cdots & Cov(V[J,B],U[J,B])
\end{array}
\right].
\end{align*}
\normalsize

For $a_1,a_2=0,1,\cdots,J$ and $b_1,b_2=0,1,\cdots,B$, applying the rectangular window function $w[i]=1,~\text{for}~i=0,\cdots,M-1$, we have
\small
\begin{align*}
&Cov(U[a_1,b_1],U[a_2,b_2]) \\
=&\sigma_n^2\sum\limits_{i_1=0}^{M-1}\sum\limits_{i_2=0}^{M-1}\delta[a_1 D+i_1-a_2D-i_2]\cos\biggl(\frac{2\pi b_1i_1}{M}\biggr)\cos\biggl(\frac{2\pi b_2i_2}{M}\biggr)
\end{align*}
\begin{align*}
&Cov(V[a_1,b_1],V[a_2,b_2]) \\
=&\sigma_n^2\sum\limits_{i_1=0}^{M-1}\sum\limits_{i_2=0}^{M-1}\delta[a_1 D+i_1-a_2D-i_2]\sin\biggl(\frac{2\pi b_1i_1}{M}\biggr)\sin\biggl(\frac{2\pi b_2i_2}{M}\biggr)
\end{align*}
\begin{align*}
&Cov(U[a_1,b_1],V[a_2,b_2]) \\
=&-\sigma_n^2\sum\limits_{i_1=0}^{M-1}\sum\limits_{i_2=0}^{M-1}\delta[a_1 D+i_1-a_2D-i_2]\cos\biggl(\frac{2\pi b_1i_1}{M}\biggr)\sin\biggl(\frac{2\pi b_2i_2}{M}\biggr)
\end{align*}
\begin{align*}
&Cov(V[a_1,b_1],U[a_2,b_2]) \\
=&-\sigma_n^2\sum\limits_{i_1=0}^{M-1}\sum\limits_{i_2=0}^{M-1}\delta[a_1 D+i_1-a_2D-i_2]\sin\biggl(\frac{2\pi b_1i_1}{M}\biggr)\cos\biggl(\frac{2\pi b_2i_2}{M}\biggr).
\end{align*}
\normalsize

When the covariance matrix is singular, we apply spectral decomposition to the covariance matrix
\begin{align*}
\bold{C_x}=Q\Lambda Q^T
\end{align*}
where $Q$ is a unitary orthogonal matrix, and $\Lambda$ is a diagonal matrix. The diagonal elements in $\Lambda$ are all real-valued. Suppose there are $k$ non-zero eigenvalues in $\Lambda$, then the covariance matrix $\bold{C_x}$ is:
\begin{align*}
\bold{C_x}=&Q\Lambda Q^{T}\\
=&\left[Q_1~~ Q_2\right]\left[\begin{array}{cc}
\Lambda_1 &0 \\
0 &\Lambda_2 \end{array}\right]\left[\begin{array}{c}
Q_1^T \\
Q_2^T \end{array}\right] \\
=&Q_1\Lambda_1Q_1^T+Q_2\Lambda_2Q_2^T
\end{align*}
\normalsize
where $\Lambda_1$ and $Q_1$ corresponds to the $k$ non-zero eigenvalues components, and $\Lambda_2$ and $Q_2$ corresponds to the zero components.

When $\bold{C_x}$ is singular, the probability density function for $\bold{X}$ does not exist. Mapping $\bold{X}$ into a subspace formed by $Q_1$, the probability density function for $Q_1^T\bold{X}$ is
\begin{align}
p(Q_1^T\bold{X})=&\frac{1}{(2\pi)^{(B+1) (J+1)}\det^{1/2}(\Lambda_1)} \notag \\
&\times \exp\biggl(-\frac{1}{2}\bold{X^T}Q_1\Lambda_1^{-1}Q_1^T\bold{X}\biggr) \label{eq:stft_h0} .
\end{align}
\normalsize

Under the $H_1$ hypothesis, letting
\begin{align*}
\bold{\mu}=&\mathbb{E}(\bold{X}),
\end{align*}
we know that
\begin{align*}
\bold{\mu}=&[\sum\limits_{i=0}^{M-1}y[0\cdot D+i]\cos(2\pi 0\cdot i/M),\ldots,\\
& \sum\limits_{i=0}^{M-1}y[J\cdot D+i]\cos(2\pi(M-1)\cdot i/M),\\
&\sum\limits_{i=0}^{M-1}y[0\cdot D+i]\sin(-2\pi 0\cdot i/M),\ldots,
\\
& \sum\limits_{i=0}^{M-1}y[J\cdot D+i]\sin(-2\pi (M-1)\cdot i/M)]^T
\end{align*}
\normalsize
and,
\vspace{-8pt}
\begin{align*}
\mathbb{E}(Q_1^T\bold{X})=Q_1^T\bold{\mu}.
\end{align*}
\normalsize

It can be proved that the covariance matrix $\bold{C_x}$ under the $H_1$ hypothesis is the same as the one under the $H_0$ hypothesis. Therefore, the probability density function under the $H_1$ hypothesis for $Q_1^T\bold{X}$ is
\begin{align}
p(Q_1^T\bold{X})=&\frac{1}{(2\pi)^{(B+1) (J+1)}\det^{1/2}(\Lambda_1)} \notag\\
&\times \exp\biggl(-\frac{1}{2}\bold{(X-\mu)^T} Q_1\Lambda_1^{-1}Q_1^T\bold{(X-\mu)}\biggr). \label{eq:stft_h1}
\end{align}
\normalsize

Based on eq.~(\ref{eq:stft_h0}) and eq.~(\ref{eq:stft_h1}), we find the likelihood ratio based on STFT data, when the environmental parameters and source signal is known exactly, to be
\begin{align}
\begin{split}
\lambda=&\frac{p(Q_1^T\bold{X}|H_1)}{p(Q_1^T\bold{X}|H_0)} \\
=&\exp\biggl(-\frac{1}{2}(\bold{\mu^T} Q_1\Lambda_1^{-1}Q_1^T\bold{\mu}-2\bold{X^T}Q_1\Lambda_1^{-1}Q_1^T\bold{\mu})
\biggr).
\end{split}
\end{align}

Therefore, the log likelihood ratio is
\begin{align}
\ln\lambda
=&-\frac{1}{2}\biggl(\bold{\mu^T} Q_1\Lambda_1^{-1}Q_1^T\bold{\mu}-2\bold{X^T}Q_1\Lambda_1^{-1}Q_1^T\bold{\mu}\biggr).
\label{eq:stft_ske_reduce1}
\end{align}
\normalsize

Because $\bold{X}$ follows a multivariate normal distribution under both hypotheses, $Q_1^T\bold{X}$  also follows a multivariate normal distribution. We can derive the log likelihood distribution under both hypothesis, and derive the analytic solution for the ROC plots:
\begin{align*}
\mu_1=&\mathbb{E}(\ln\lambda|H_1)=\frac{1}{2}(\bold{\mu^T}Q_1\Lambda_1^{-1}Q_1^T\bold{\mu}) \\
\sigma_1^2=&Var(\ln\lambda|H_1) \notag\\
=&(Q_1\Lambda_1^{-1}Q_1^T\bold{\mu})^TCov(\bold{X})Q_1\Lambda_1^{-1}Q_1^T\bold{\mu} \notag\\
=&(Q_1\Lambda_1^{-1}Q_1^T\bold{\mu})^T(Q_1\Lambda_1Q_1^T+Q_2\Lambda_2Q_2^T)Q_1\Lambda_1^{-1}Q_1^T\bold{\mu}
\notag\\
=&\bold{\mu^T}Q_1\Lambda_1^{-1}Q_1^T\bold{\mu} \\
\mu_0=&\mathbb{E}(\ln\lambda|H_0)=-\frac{1}{2}(\bold{\mu^T}Q_1\Lambda_1^{-1}Q_1^T\bold{\mu}) \\
\sigma_0^2=&Var(\ln\lambda|H_0)=Var(\ln\lambda|H_1)=\bold{\mu^T}Q_1\Lambda_1^{-1}Q_1^T\bold{\mu}.
\end{align*}
\normalsize

Therefore,
\begin{align}
\ln\lambda|H_1&\sim\mathcal{N}\Bigl(\frac{1}{2}(\bold{\mu^T}Q_1\Lambda_1^{-1}Q_1^T\bold{\mu}),~\bold{\mu^T}Q_1\Lambda_1^{-1}Q_1^T\bold{\mu}\Bigr) \label{eq:log_likelihood_h1_ske} \\
\ln\lambda|H_0&\sim\mathcal{N}\Bigl(-\frac{1}{2}(\bold{\mu^T}Q_1\Lambda_1^{-1}Q_1^T\bold{\mu}),~\bold{\mu^T}Q_1\Lambda_1^{-1}Q_1^T\bold{\mu}\Bigr) . \label{eq:log_likelihood_h0_ske}
\end{align}
\normalsize

The probability of correct detection $P_D$ and the probability of false alarm $P_F$ in this case are~\cite{StevenKay}:
\begin{align*}
P_D&=\int^{\infty}_{\ln\beta}p(\ln\lambda|H_1)d(\ln\lambda)\\
&=\frac{1}{\sqrt{2\pi\sigma_1^2}}\int^{\infty}_{\ln\beta}e^{-\frac{(\ln\lambda-\mu_1)^2}{2\sigma_1^2}}d(\ln\lambda)\\
P_F&=\int^{\infty}_{\ln\beta}p(\ln\lambda|H_0)d(\ln\lambda)\\
&=\frac{1}{\sqrt{2\pi\sigma_0^2}}\int^{\infty}_{\ln\beta}e^{-\frac{(\ln\lambda-\mu_0)^2}{2\sigma_0^2}}d(\ln\lambda).
\end{align*}
\normalsize

Since $\sigma_1^2=\sigma_0^2$, the standard deviation in this case is positive, so $\sigma_1=\sigma_0$. Letting $\sigma=\sigma_1=\sigma_0$, $z=\frac{(\ln\lambda-\mu_1)}{\sigma}$, $z'=\frac{(\ln\lambda-\mu_0)}{\sigma}$, $\sigma'=(\ln\beta-\mu_0)/\sigma$, so $dz=\sigma d(\ln\lambda)$, $dz'=\sigma d(\ln\lambda)$, we have
\begin{align}
P_F&=\frac{1}{\sqrt{2\pi}}\int^{\infty}_{\sigma'}e^{-\frac{z'^2}{2}}dz' ,\\
P_D&=\frac{1}{\sqrt{2\pi}}\int^{\infty}_{\sigma'-\frac{\mu_1-\mu_0}{\sigma}}e^{-\frac{z^2}{2}}dz,
\end{align}
\normalsize
and the separation parameter, or detection index, $d$ in this case is
\begin{align*}
d&=\frac{\mu_1-\mu_0}{\sigma},
\end{align*}
\normalsize
where,
\begin{align}
d^2&=\frac{(\mu_1-\mu_0)^2}{(\sigma)^2} \notag\\
&=\frac{(\bold{\mu^T}Q_1\Lambda_1^{-1}Q_1^T\bold{\mu} )^2}{\bold{\mu^T}Q_1\Lambda_1^{-1}Q_1^T\bold{\mu}}=\bold{\mu^T}Q_1\Lambda_1^{-1}Q_1^T\bold{\mu}.
\end{align}
\normalsize

\medskip

\subsection{\textbf{Mean ocean}}
The following is a proof that when the uncertainty of the sound speed profile is small, we can theoretically approximate the likelihood ratio using eq.~(\ref{eq:lambda_stft_mean}). Suppose $p_k$ is the probability of the sound speed profile of the $k^{th}$ path; $\bold{\mu_k}$ is the propagated signal corresponding to the $k^{th}$ sound speed profile, and $\bold{\mu_m}$ is the propagated signal of the mean sound speed profile. According to eq.~(\ref{eq:signal_prop}), we can see that the mapping of the sound speed and propagated signal is linear, so we have $\bold{\mu_m}=\sum\limits_{k=1}^{P}p_k\bold{\mu_k}$. According to eq.~(\ref{eq:lambda_stft}), the geometric mean of the likelihood ratio is
\small
\begin{align*}
&\prod_{k=1}^{P}\lambda_k^{p_k} \notag \\
=&\exp\Biggl(-\frac{1}{2}\biggl(\sum\limits_{k=1}^{P}p_k\Bigl(\bold{\mu_k^T} Q_1\Lambda_1^{-1}Q_1^T\bold{\mu_k}-2\bold{X^T}Q_1\Lambda_1^{-1}Q_1^T\bold{\mu_k}\Bigr)\biggr)
\Biggr) \notag \\
=&\exp\Biggl(\biggl(-\frac{1}{2}\sum\limits_{k=1}^{P}p_k\bold{\mu_k^T} Q_1\Lambda_1^{-1}Q_1^T\bold{\mu_k}\biggr)+\bold{X^T}Q_1\Lambda_1^{-1}Q_1^T\bold{\mu_m}
\Biggr)
\end{align*}
\normalsize

For the term $-\frac{1}{2}\sum\limits_{k=1}^{P}p_k\bold{\mu_k^T} Q_1\Lambda_1^{-1}Q_1^T\bold{\mu_k}$, notice that:
\begin{align*}
&\sum\limits_{k=1}^{P}p_k\bold{\mu_k^T} Q_1\Lambda_1^{-1}Q_1^T\bold{\mu_k} \\
=&\sum\limits_{k=1}^{P}p_k\bold{\mu_k^T} Q_1\Lambda_1^{-1}Q_1^T(\bold{\mu_k-\mu_m}) \\
=&\sum\limits_{k=1}^{P}p_k(\bold{\mu_k^T-\mu_m^T}) Q_1\Lambda_1^{-1}Q_1^T(\bold{\mu_k-\mu_m}) \\
&+\sum\limits_{k=1}^{P}p_k\bold{\mu_m^T} Q_1\Lambda_1^{-1}Q_1^T(\bold{\mu_k-\mu_m}) \\
=&\sum\limits_{k=1}^{P}p_k(\bold{\mu_k^T-\mu_m^T}) Q_1\Lambda_1^{-1}Q_1^T(\bold{\mu_k-\mu_m}) \\
\leq & C\sum\limits_{k=1}^{P}p_k||\bold{\mu_k-\mu_m}||^2 = C Var(\bold{\mu_k})
\end{align*}
\normalsize
where $C$ is the largest eigenvalue of $\Lambda_1^{-1}$. When the variance of $\mu_k$ is small, we can approximate $\sum\limits_{k=1}^{P}p_k\bold{\mu_k^T} Q_1\Lambda_1^{-1}Q_1^T\bold{\mu_k}$ with $\sum\limits_{k=1}^{P}p_k\bold{\mu_m^T} Q_1\Lambda_1^{-1}Q_1^T\bold{\mu_m}$. Therefore,
\small
\begin{align*}
&\prod_{k=1}^{P}\lambda_k^{p_k}\approx \exp\Biggl(-\frac{1}{2}\biggl(\bold{\mu_m^T} Q_1\Lambda_1^{-1}Q_1^T\bold{\mu_m}-2\bold{X^T}Q_1\Lambda_1^{-1}Q_1^T\bold{\mu_m}\biggr)
\Biggr).
\end{align*}
\normalsize

We can see that $\bold{\mu_m}$ in eq.~(\ref{eq:lambda_stft_mean}) is the geometric mean of the likelihood ratio.

\medskip

\section*{Acknowledgment}
 The authors would like to thank Dr.~Surya Tokdar, Dr.~Andrew Thompson, Dr.~Granger Hickman, Dr.~Yingbo Li, Dr.~Galen Reeves for helpful discussion and instructions.

\medskip

\ifCLASSOPTIONcaptionsoff
  \newpage
\fi

%





\end{document}